\documentclass[smallcondensed]{svjour3}

\usepackage[english]{babel}
\usepackage[utf8]{inputenc}
\usepackage{amsmath}
\usepackage{amsfonts}
\usepackage{graphicx}
\usepackage{ifoddpage}
\usepackage[margin=1in]{geometry}
\usepackage{setspace}
\usepackage{xcolor}

\usepackage{csquotes}
\usepackage{authblk}
\usepackage{appendix}
\journalname{Quantum Information Processing}

\newtheorem{lem}{Lemma}

\title{Qutrit-based semi-quantum key distribution protocol}

\author{Hasnaa Hajji \and Morad El Baz}
\institute{ESMaR, Faculty of Sciences, Mohammed V University in Rabat, Morocco. \email{hasnaa$\_$hajji@um5.ac.ma \and morad.elbaz@um5.ac.ma}}

\begin{document}

\maketitle

\begin{abstract}

\par This article provides the unconditional security of a semi quantum key distribution (SQKD) protocol based on 3-dimensional quantum states. By deriving a lower bound for the key rate, in the asymptotic scenario, as a function of the quantum channel's noise, we find that this protocol has improved secret key rate with much more tolerance for noise compared to the previous 2-dimensional SQKD protocol. Our results highlight that, similar to the fully quantum key distribution protocol, increasing the dimension of the system can increase the noise tolerance in the semi-quantum key distribution, as well.

\end{abstract}
\section {Introduction}

 In 2007 Boyer \cite{1} suggested a semi quantum key distribution protocol (SQKD) which enables two parties to establish a secret and secure key using principles of quantum mechanics. In this scheme, Alice, the sender of the quantum information, transmits quantum bits in the computation basis $\{|0\rangle, |1\rangle\}$ or the conjugate basis, $\{|\pm\rangle=1/\sqrt{2}(|0\rangle\pm |1\rangle \}$ to, Bob, the receiver, who is restricted to performing one of two operations:
  \begin{itemize}
  \item\textsc{Measure and Resend:} measuring the received qubits in the computation basis and resending his result to Alice. (i.e Bob will generate the state he measured and resends it towards Alice)
    
    \item\textsc{Reflect:} Ignoring the quantum bit and reflecting it back to Alice, learning nothing about its state in the process.
  \end{itemize}
  
 Since Bob is limited to the above operations, he is typically referred to as a classical user.
 
 SQKD protocols require a two-way quantum communication channel, which allows a quantum bit to travel from Alice to Bob and then back. This gives, Eve, the attacker, two opportunities to interact with the quantum bit, thus substantially increasing the complexity of the security analysis. For this reason, up until 2015, most security proofs for SQKD protocols focused on the notion of \textit{robustness}(see \cite{1},\cite{2},\cite{3},\cite{4},\cite{5},\cite{6},\cite{7},\cite{10}). In 2015, Krawec \cite{2U} gave the proof of the unconditional security of Boyer's original SQKD protocol. Then, in 2017, Krawec \cite{1U} also proved that a SQKD protocol can tolerate the same noise level as the fully quantum BB84 protocol. Besides, in 2016 Zhang \textit{et al.} demonstrated the unconditional security of the SQKD protocol using less than four quantum states\cite{3U}.
 
Due to the interesting features of the semi-quantum protocols, researchers have shown great enthusiasm on it and absorbed it into traditional quantum cryptography tasks such as secret sharing \cite{1S,2S,3S}, private comparison \cite{1P,2P,3P}, and direct communication \cite{1DC,2DC,3DC}. All these works have been carried out based on two level systems: \textit{qubits}, as the fundamental unit of quantum information.

Independently of this, high dimensional systems (\textit{qudits}) have attracted much attention with many theoretical proposals and experimental realizations \cite{peresbook}. They offer advantages ranging from withstanding high channel noise levels, to possible implementation in the fundamental tests of quantum mechanics. As a result of the dimension increase, quantum computing and quantum information processing in general, gets more efficient with qudits, as one can achieve the same computational dimension employing fewer qudits \cite{sandersqudit}. High dimensional quantum key distribution was introduced \cite{PhysRevLett.85.3313}, realized experimentally \cite{experimentalqudit} and was shown to increase the channel capacity and thus the key generation rate \cite{highdimqkd}. This paper introduces a generalized scheme of the SQKD protocol from \cite{1} based on the use of three-dimensional quantum states: \textit{qutrits}. The use of qurtits, the simplest system capable of exhibiting quantum contextuality \cite{peresbook}, has been proven to carry more information and have greater resistance to noise than qubits. Encouraged by this, we use the technique developed in \cite{2U} to produce an expression for the key rate of this protocol, in the asymptotic scenario, as a function of the quantum channel's noise.

The rest of the paper is organized as follows: the next section, introduces some preliminaries about SQKD. Section 3 describes the SQKD protocol using qutrits as information carriers. Section 4 presents a security analysis of our proposal while results and discussions are given in section 5. A conclusion with some perspectives are given at the end of the paper.

\section{Preliminaries}

In this section we present some notations and definitions that will be used throughout the paper.

\subsection{Entropy of the shared states:}

Let $\rho$ be a density operator, i.e. a Hermitian positive semi-definite operator of unit trace, acting on some finite dimensional Hilbert space $\mathcal{H}$. If $\rho$ is acting on a bipartite space $\mathcal{H}_{A} \otimes\mathcal{H}_{B}$, we often write it as $\rho_{AB}$; in this case, $\rho_{B}$ means the reduced density matrix resulting from tracing out over subsystem A \textit{i.e.} the partial trace: $\rho_{B}= tr_{A}\rho_{AB}$.

$S(AB)$ denotes the von Neumann entropy of $\rho_{AB}$:
\begin{equation}
    S(AB)= S(\rho_{AB})=-tr\left( \rho_{AB}\log_3{\rho_{AB}}\right)=-\sum_{i} \lambda_{i}\log_{3}\lambda_{i}
\end{equation} 
where the ${\lambda_{i}}$'s are the eigenvalues of $\rho_{AB}$. Note that, since our protocol is based on qutrits, we use the base three logarithm instead of base two, as is customary the case, when dealing with qubits. Similarly, $S(B)$ denotes the von Neumann entropy of $\rho_{B}$. We also write $S(A|B)$ to mean the conditional von Neumann entropy \cite{nielsen2002quantum}, namely 
\begin{eqnarray}
  S(A|B)= S(AB)-S(B)= S(\rho_{AB})-S(\rho_{B}).  
\end{eqnarray}

We use $H(p_{1},p_{2},\dots ,p_{n})$ to represent the classical Shannon entropy of the probability distribution $\left\{p_{1},p_{2},\dots,p_{n}\right\}$: $H(p_{1},p_{2},\dots,p_{n})=-\sum_{i} p_{i} \log p_{i}$. Unless otherwise specified all logarithms in this paper are base three.
\begin{lem}
\label{lemma1}
Given a finite dimensional Hilbert space $\mathcal{H}=\mathcal{H}_C\otimes\mathcal{H}_E$, let $\{ |1 \rangle_C,\dots, |n \rangle_C \}$ be an orthonormal basis of $\mathcal{H}_C$. Consider the following density operator acting on $\mathcal{H}$:
\begin{equation}
    \rho =\sum^{n}_{j=1} p_j |j \rangle\langle j|_C\otimes\sigma_E^{(j)},
\end{equation}
where $\sum p_i=1,p_i\geq 0$, and each $\sigma^{(i)}_E$ is a Hermitian positive semi-definite operator of unit trace acting on $\mathcal{H}_E$. Then the von Neumann entropy of $\rho$ is
\begin{equation}
    S(\rho)=H(p_{1},p_{2},\dots,p_{n})+\sum^{n}_{j=1}p_j S(\sigma^{(i)}_E).
\end{equation}
\end{lem}

For a proof of this Lemma, one may refer to the book \cite{nielsen2002quantum}.

\subsection{Qutrit bases:}

Let $\{|0\rangle, |1\rangle, |2\rangle \}$ be the unit vectors of the basis $\mathcal{A}$ of a three dimensional Hilbert space (the state space of a qutrit system). Alternative choices of bases would be the $\mathcal{T}$ basis, given by vectors $\{|0^{\prime}\rangle, |1^{\prime}\rangle, |2^{\prime}\rangle\}$:
\begin{align}
\label{Tbasis}
    |0^\prime \rangle &=\frac{1}{\sqrt{3}}(e^{\frac{2i\pi}{3}}|0\rangle+ |1\rangle+ |2\rangle),\nonumber \\ 
    |1^\prime \rangle &=\frac{1}{\sqrt{3}}(|0\rangle+e^{\frac{2i\pi}{3}}|1\rangle+ |2\rangle),\\
      |2^\prime \rangle &=\frac{1}{\sqrt{3}}(|0\rangle+ |1\rangle+e^{\frac{2i\pi}{3}}|2\rangle),\nonumber
\end{align}
and the $\mathcal{K}$ basis, given by vectors $\{|0^{\prime\prime}\rangle, |1^{\prime\prime}\rangle, |2^{\prime\prime}\rangle\}$:
\begin{align}
\label{Kbasis}
    |0^{\prime\prime} \rangle &=\frac{1}{\sqrt{3}}(|0\rangle+ |1\rangle+ |2\rangle),\nonumber \\ 
    |1^{\prime\prime} \rangle &=\frac{1}{\sqrt{3}}(|0\rangle+e^{\frac{2i\pi}{3}}|1\rangle+e^{\frac{-2i\pi}{3}} |2\rangle),\\
    |2^{\prime\prime} \rangle &=\frac{1}{\sqrt{3}}(|0\rangle+e^{\frac{-2i\pi}{3}} |1\rangle+e^{\frac{2i\pi}{3}}|2\rangle).\nonumber
\end{align}

It is worthwhile noting that other choices of bases are possible \cite{PhysRevLett.85.3313}, however we restrict ourselves to these two bases as they are the natural choice when extending the usual two mutually unbiased bases in qubits based protocols such as BB84 \cite{bennett2020quantum} or the semi-quantum ones \cite{1}. It turns out (see section \ref{sect-security}) that these two choices are those leading to optimal results in terms of security or key rate.

\subsection{(S)QKD security}

Any key distribution protocol, being quantum or semi-quantum, operates in two stages. The first, called the quantum communication stage, performs several iterations to agree on a so-called raw key of size N bits. Following this, a classical stage consisting of error correction and privacy amplification is run, resulting in a secret key of size $l(N)\leq N$ (see \cite{RevModPhys.81.1301} for more information on these, now standard, processes).  The key rate, $r$, in the asymptotic scenario is defined as \cite{8,Renner_2005}
\begin{equation}
    r=\lim_{N\rightarrow \infty}\frac{l(N)}{N}.
\end{equation}

\begin{figure}
    \centering
   
    \includegraphics[scale=0.5]{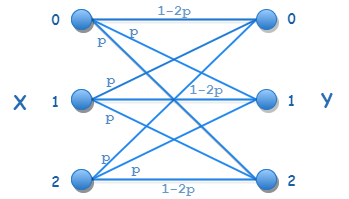}
    \caption{A symmetric ternary channel, where a trit $i$ from the input $X$ has a probability $p$ to be flipped into trit $j$ in the output $Y$ ($i\neq j = {0,1,2}$).}
    \label{tcfig}
\end{figure}

Based on the observed parameters, we wish to evaluate the key rate expression under a particular attack scenario. Namely, we will consider a symmetric attack modeled by a ternary channel as depicted in Fig. \ref{tcfig}. A noisy quantum channel modeled as a ternary channel with parameter Q is simply the map :
\begin{equation}
    \xi(\rho)=(1-3Q)\rho +QI,
\end{equation}
where $I$ is the identity operator.

\section{The Protocol}

The protocol we consider in this paper is an extension of the qubit based semi-quantum key distribution (SQKD) protocol introduced in \cite{1} to a  three dimensional case (qutrits). Two natural choices (\ref{Tbasis}, \ref{Kbasis}) for the two bases that Alice is allowed to use emerge as a result of working with qutrits instead of qubits; we study both versions of the protocol.

 Let $\Phi_1$  (respectively $\Phi_2$) be the set of six states in the bases $\mathcal{A}$ and $\mathcal{T}$ (respectively $\mathcal{A}$ and $\mathcal{K}$) from which Alice may choose the state to send to Bob:
\begin{eqnarray}
    \Phi_1 &=&\left\{|0\rangle, |1\rangle, |2\rangle, |0^{\prime}\rangle, |1^{\prime}\rangle, |2^{\prime}\rangle\right\}, \\
     \Phi_2 &=&\left\{|0\rangle, |1\rangle, |2\rangle,|0^{\prime\prime}\rangle, |1^{\prime\prime}\rangle, |2^{\prime\prime}\rangle\right\}.
\end{eqnarray}

The steps of the protocol being the same, independently of the choice of bases, we present the protocol in what follows and we delay the distinction between the two cases, to the discussion of the security  and the key rate in section \ref{sect-security}, for which the choice of bases is crucial.

\begin{figure}[h]
    \centering
 
    \includegraphics[scale=0.25]{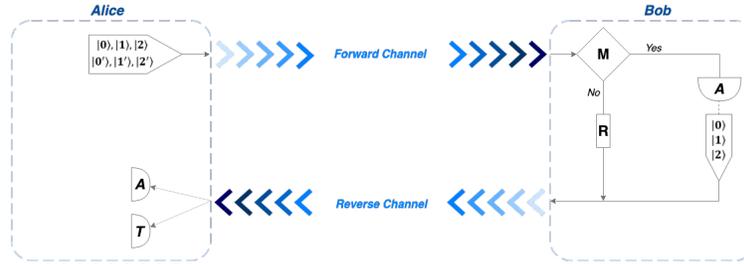}
    \caption{3-dimensional SQKD protocol: The fully quantum user, Alice, prepares one of six states from the set $\phi_1$,sends it to the classical user, Bob, through the forward channel. Bob either ignores the qutrit and returns it to Alice (\textbf{R}) or performs a measurement on the basis $\mathcal{A}$ (\textbf{M}). Through the reverse channel, He sends the measured state to Alice who measures it in the same basis she prepared it initially ($\mathcal{A}$ or $\mathcal{T}$ basis)}
\end{figure}

\textbf{Quantum Communication Stage}: The quantum communication stage of the protocol repeats the following single iteration until a sufficiently large raw key is established:
\begin{enumerate}
    \item Alice, the fully quantum capable user, randomly prepares a qutrit in a state from  the set $\Phi_i$, where $i=1,2$ depending on the version of the protocol adopted. This qutrit is then sent to Bob.
    
    \item Bob, being the classical user, is restricted to randomly choosing to either reflect the qutrit ($R$) or to measure it in the basis $\mathcal{A}$ and then re-sending it in the measured state \footnote{Bob's qualification as a classical user stems from his inability to use (measure and/or prepare states in) more than one basis, unlike Alice, the fully quantum user, capable of using both the $\mathcal{A}$ and $\mathcal{K}$ bases (or the $\mathcal{A}$ and $\mathcal{T}$ bases, based on the version of the protocol adopted).} (henceforward denoted operation $M$). In this last case, Bob saves the result of his measurement as his potential raw key. However he does not, yet, reveal his choice of operation ($R$ or $M$).

    \item Upon receiving her qutrit back from Bob, Alice will measure it in the same basis she originally used to prepare it in, in step 1.

\textbf{Classical Communication Stage}:
 \item Using an authenticated classical channel, Alice reveals her choice of basis while Bob reveals his choice of operation ($M$ or $R$).
 
    \begin{itemize}
         \item  If Alice chose the $\mathcal{A}$ basis to encode her qutrit and Bob chose to measure and resend it $(M)$, they keep their respective results to form their respective version of the raw key.
         
         \item Using a randomly chosen sample of this raw key, Alice and Bob divulge their measurement results to estimate the noise level in the channel.
         They may also observe the noise in the $\mathcal{K}$ or $\mathcal{T}$ (depending on the set chosen for the protocol $\Phi_1$ or $\Phi_2$) using those iterations where Alice chose to encode her qutrit in the $\mathcal{K}$ or a $\mathcal{T}$ basis, respectively, while Bob chose to reflect it.  
    \end{itemize}

\item Assuming the channel noise is low enough, Alice and Bob perform error correction and privacy amplification, resulting in a secret key. 
\end{enumerate}

\section{Security proof against collective attacks}
\label{sect-security}
  The security of (S)QKD protocols against sophisticated attacks is among the most important issues in quantum information theory. In this section we prove the security of our protocol against a very important class of attacks called collective attacks. In conjunction with that, we bound the von Neumann entropy and the key rate. The final results depend on the choice of the alternative basis Alice uses to encode her qubits, the $\mathcal{K}$ or $\mathcal{T}$ basis. Both cases will be treated and compared.
  
\subsection{Collective attacks}

 In collective attacks Eve keeps her ancillary states (or probes) in a quantum memory till receiving all classical data including error-correction and privacy amplification data, and performs the optimal measurement on her ancillary states (or probes) to learn the maximal information on the final key.

Assuming collective attacks, we may employ the Devetak-Winter \cite{8} key rate equation: 
\begin{equation}
\label{keyrateinf}
    r=inf[S(B|E)-H(B|A)],
\end{equation}
which states that the key rate is the difference between Eve's uncertainty on Bob's raw key, $S(B|E)$ (which, for a successful protocol,  should be high) and Alice's uncertainty of Bob's key, $H(B|A)$ (which should be low). The infimum is over all collective attacks that induce the observed error rates. From this, our main focus in this work is to determine a lower bound on our protocol's key rate. To do so, we will first describe the quantum system after one successful iteration of the protocol.

As is generally the case with the security analysis of any (S)QKD protocol, we assume that Eve is powerful enough to control the natural noise and without loss of generality, we may assume that her ancilla is in some initial state $|0\rangle_E$. Eve's most general attack comprises two unitary operations: $U_F$ attacking qutrits as they go from Alice to Bob (the forward direction) and $U_R$ as they go back from Bob to Alice (the reverse direction).

So in the collective attack, the state $|a\rangle \otimes |0\rangle_E $, where $|a\rangle$ is the qutrit' state sent by Alice to Bob, will be subjected, in the forward channel, to Eve's unitary transformation operators $U_F$ and will yield
\begin{eqnarray}
\label{UF}
|0\rangle\otimes |0\rangle_{E} &\xrightarrow{U_F}&|0,e_0\rangle +|1,e_1\rangle+|2,e_2\rangle,  \nonumber \\
|1\rangle\otimes |0\rangle_{E} &\xrightarrow{U_F}& |0,e_3\rangle+|1,e_4\rangle+|2,e_5\rangle , \label{uf1} \\
|2\rangle\otimes |0\rangle_{E} &\xrightarrow{U_F}& |0,e_6\rangle+|1,e_7\rangle+|2,e_8\rangle . \nonumber
\end{eqnarray}
On the backward channel, Bob's qutrit state will be altered by the operation $U_R$ and will yield the final global state as follows

\begin{equation}
\label{ur}
    |i, e_j\rangle \xrightarrow{U_R} |0, e^0_{i,j} \rangle + |1, e^1_{i,j}\rangle+ |2, e^2_{i,j}\rangle,
\end{equation}
where $i=0, 1, 2$, $j=0, 1, \dots ,8$. The vectors  $|e_i\rangle$ and $ |e^k_{i,j} \rangle $ are Eve's ancillary states, not necessarily normalized nor orthogonal, after the attack on the forward channel and reverse channel respectively. However, due to the unitarity of $U_F$ and $ U_R$ we have
\begin{equation}
    \langle e_{0} |e_{0} \rangle +\langle e_{1} |e_{1} \rangle +\langle e_{2} |e_{2} \rangle=\langle e_{3} |e_{3} \rangle +\langle e_{4} |e_{4} \rangle +\langle e_{5} |e_{5} \rangle=\langle e_{6} |e_{6} \rangle +\langle e_{7} |e_{7} \rangle +\langle e_{8} |e_{8} \rangle=1,
\end{equation}
\begin{equation}
   \langle e_{0} |e_{3} \rangle +\langle e_{1} |e_{4} \rangle +\langle e_{2} |e_{5} \rangle=\langle e_{3} |e_{6} \rangle +\langle e_{4} |e_{7} \rangle +\langle e_{5} |e_{8} \rangle=\langle e_{0} |e_{6} \rangle +\langle e_{1} |e_{7} \rangle +\langle e_{2} |e_{8} \rangle=0.
\end{equation}

\par Following Eve's attack, and conditioning on an iteration being used to contribute towards the raw key (\textit{i.e} Alice sent $|0\rangle$ , $|1\rangle$ or $|2\rangle$ with probability 1/3 each and Bob measured it and resent it), the density operator describing Bob and Eve's system after Alice's measurement is
\begin{align}
\label{rhobe}
     \rho_{BE} & = \frac{1}{3}|0\rangle \langle0|_B\otimes \Big( |e^0_{0,0}\rangle \langle e^0_{0,0}|+ | e^1_{0,0} \rangle \langle  e^1_{0,0}  |+ | e^2_{0,0} \rangle \langle  e^2_{0,0}  |+ |e^0_{0,3}\rangle \langle e^0_{0,3}| \nonumber \\
     & +| e^1_{0,3} \rangle \langle  e^1_{0,3}  |+| e^2_{0,3} \rangle \langle  e^2_{0,3}  |+|e^0_{0,6}\rangle \langle e^0_{0,6}|+ | e^1_{0,6} \rangle \langle  e^1_{0,6}|+ | e^2_{0,6} \rangle \langle  e^2_{0,6}| \Big) \nonumber \\
     & + \frac{1}{3} |1\rangle \langle1|_B\otimes\Big(|e^0_{1,1}\rangle \langle e^0_{1,1}|+ | e^1_{1,1} \rangle \langle  e^1_{1,1}  |+ | e^2_{1,1} \rangle \langle  e^2_{1,1}  |+ |e^0_{1,4}\rangle \langle e^0_{1,4}|\nonumber \\
     & + | e^1_{1,4} \rangle \langle  e^1_{1,4}|+ | e^2_{1,4} \rangle \langle  e^2_{1,4}|+ |e^0_{1,7}\rangle \langle e^0_{1,7}|+ | e^1_{1,7} \rangle \langle  e^1_{1,7}|+ | e^2_{1,7} \rangle \langle  e^2_{1,7}|\Big)\nonumber \\ 
     & + \frac{1}{3} |2\rangle \langle2|_B\otimes\Big(|e^0_{2,2}\rangle \langle e^0_{2,2}|+ | e^1_{2,2} \rangle \langle  e^1_{2,2}  |+ | e^2_{2,2} \rangle \langle  e^2_{2,2}|+ |e^0_{2,5}\rangle \langle e^0_{2,5}|\nonumber \\ 
     & + | e^1_{2,5} \rangle \langle  e^1_{2,5}|+ | e^2_{2,5} \rangle \langle  e^2_{2,5}|+ |e^0_{2,8}\rangle \langle e^0_{2,8}|+ | e^1_{2,8} \rangle \langle  e^1_{2,8}|+ | e^2_{2,8} \rangle \langle  e^2_{2,8}|\Big).
\end{align}

This state will be used next, to bound the von Neumann entropy in equation (\ref{keyrateinf}).

\subsection{Bounding the von Neumann entropy}
 Our goal in the remainder of the security proof is to bound the von Neumann Entropy $S(B|E)$. However, due to the higher dimension of the system, this might prove to be a difficult task. To overcome this, we will employ a technique proposed in \cite{christ2004generic} and also used in \cite{2U}. This technique requires to condition on an additional random variable $C$, in order to simplify the entropy computations. Owing to the strong sub-additivity of the von Neumann entropy, for any tripartite system, it holds that
 \begin{equation}
 \label{strongsubadditivity}
 S(B|E)\geq S(B|EC) \implies S(B|E)-H(B|A)\geq S(B|EC)-H(B|A),
\end{equation}
thus supplying us with a lower bound on the key rate of this protocol. 

The system, $c$, we append will be in a four-dimensional space, spanned by $\{|c,0\rangle, |c,1\rangle, |w,1\rangle, |w,2\rangle \}$ where $|c,i \rangle \langle c, i|$ is the event that Alice and Bob's raw key bits match (\textit{i.e.} are correct), and that the qutrit sent from Alice was flipped $i$ times, while $|w,i \rangle \langle w, i|$ denotes the event where their raw key bits don't match (\textit{i.e.} are wrong). 

Incorporating this system, we can rewrite the global state system (\ref{rhobe}) as
\begin{align}
     \rho_{BEC} & = \frac{1}{3}|0\rangle \langle0|_B\otimes(|e^0_{0,0}\rangle \langle e^0_{0,0}|\otimes|c,0 \rangle \langle c, 0|+ | e^1_{0,0} \rangle \langle  e^1_{0,0}  |\otimes|w,1 \rangle \langle w, 1|+|e^2_{0,0} \rangle \langle  e^2_{0,0}|\otimes|w,1 \rangle \langle w, 1  |\nonumber \\
     &+ |e^0_{0,3}\rangle \langle e^0_{0,3}|\otimes|c,1 \rangle \langle c, 1|+|e^1_{0,3} \rangle\langle e^1_{0,3}|\otimes |w,2 \rangle \langle w, 2|+| e^2_{0,3} \rangle \langle e^2_{0,3}| \otimes |w,2 \rangle \langle w, 2|\nonumber \\
     &+|e^0_{0,6}\rangle \langle e^0_{0,6}| \otimes |c,1 \rangle \langle c,1| + | e^1_{0,6} \rangle \langle  e^1_{0,6}|  \otimes |w,2 \rangle \langle w, 2|+| e^2_{0,6} \rangle \langle  e^2_{0,6}|\otimes |w,2 \rangle \langle w, 2|)\nonumber \\
     &+ \frac{1}{3} |1\rangle \langle1|_B\otimes(|e^0_{1,1}\rangle \langle e^0_{1,1}|\otimes |w,2 \rangle \langle w, 2| +|e^1_{1,1} \rangle \langle e^1_{1,1}|\otimes |c,1 \rangle \langle c,1| +| e^2_{1,1} \rangle \langle e^2_{1,1}|\otimes |w,2 \rangle \langle w, 2|\nonumber \\
     &+|e^0_{1,4}\rangle \langle e^0_{1,4}|\otimes |w,1 \rangle \langle w,1| + | e^1_{1,4} \rangle \langle  e^1_{1,4}| \otimes |c,0 \rangle \langle c,0| +  | e^2_{1,4} \rangle \langle  e^2_{1,4}|\otimes  |w,1 \rangle \langle w,1| \nonumber \\
     &+|e^0_{1,7}\rangle \langle e^0_{1,7}|\otimes |w,2 \rangle \langle w, 2| + | e^1_{1,7} \rangle \langle  e^1_{1,7}| \otimes |c,1 \rangle \langle c,1| +| e^2_{1,7} \rangle \langle  e^2_{1,7}|\otimes |w,2 \rangle \langle w, 2|)\nonumber \\ 
     & + \frac{1}{3} |2\rangle \langle2|_B\otimes(|e^0_{2,2}\rangle \langle e^0_{2,2}| \otimes |w,2 \rangle \langle w, 2| +| e^1_{2,2} \rangle \langle  e^1_{2,2}  | \otimes |w,2 \rangle \langle w, 2| + | e^2_{2,2} \rangle \langle  e^2_{2,2}| \otimes |c,1 \rangle \langle c, 1| \nonumber\\
     &+ |e^0_{2,5}\rangle \langle e^0_{2,5}| \otimes |w,2 \rangle \langle w, 2| + | e^1_{2,5} \rangle \langle  e^1_{2,5}| \otimes |w,2 \rangle \langle w, 2| +| e^2_{2,5} \rangle \langle  e^2_{2,5}| \otimes |c,1 \rangle \langle c,1| \nonumber \\
     &+|e^0_{2,8}\rangle \langle e^0_{2,8}| \otimes |w,1 \rangle \langle w, 1| +| e^1_{2,8} \rangle \langle  e^1_{2,8}| \otimes |w,1 \rangle \langle w, 1| + | e^2_{2,8} \rangle \langle  e^2_{2,8}| \otimes |c,0 \rangle \langle c, 0|). \label{rhobec}
\end{align}

 Let $p_{i,j,k}$ denote the probability that Alice initially sends the state $|i\rangle$, Bob finds the results $|j \rangle$ when measuring it and Alice finds $|k \rangle$ when measuring the qutrit returned by Bob. If there is no noise in the $\mathcal{A}$ basis (which is what we are going to assume for now), it should hold that $p_{0,0,0}=p_{1,1,1}=p_{2,2,2}=1$. These probabilities can be used to estimate the value $\langle e^a_{b,c}| e^a_{b,c} \rangle$ as follows:
\begin{equation*}
  p_{0,0,0} = \langle e^0_{0,0} |e^0_{0,0} \rangle \hspace{0.5cm}  p_{1,0,0} = \langle e^0_{0,3} |e^0_{0,3}\rangle \hspace{0.5cm}  p_{2,0,0} = \langle e^0_{0,6} |e^0_{0,6}\rangle
\end{equation*}
\begin{equation*}
  p_{0,0,1} = \langle e^1_{0,0} |e^1_{0,0} \rangle \hspace{0.5cm}  p_{1,0,1} = \langle e^1_{0,3} |e^1_{0,3}\rangle \hspace{0.5cm} p_{2,0,1} = \langle e^1_{0,6} |e^1_{0,6}\rangle
\end{equation*}
\begin{equation*}
  p_{0,0,2} = \langle e^2_{0,0} |e^2_{0,0} \rangle \hspace{0.5cm}  p_{1,0,2} = \langle e^2_{0,3} |e^2_{0,3}\rangle \hspace{0.5cm} p_{2,0,2} = \langle e^2_{0,6} |e^2_{0,6}\rangle
\end{equation*}
\begin{equation*}
  p_{0,1,0} = \langle e^0_{1,1} |e^0_{1,1} \rangle \hspace{0.5cm}  p_{1,1,0} = \langle e^0_{1,4} |e^0_{1,4} \rangle \hspace{0.5cm} p_{2,1,0} = \langle e^0_{1,7} |e^0_{1,7} \rangle
\end{equation*}
\begin{equation*}
  p_{0,1,1} = \langle e^1_{1,1} |e^1_{1,1} \rangle \hspace{0.5cm}  p_{1,1,1} = \langle e^1_{1,4} |e^1_{1,4} \rangle \hspace{0.5cm} p_{2,1,1} = \langle e^1_{1,7} |e^1_{1,7} \rangle
\end{equation*}
\begin{equation*}
  p_{0,1,2} = \langle e^2_{1,1} |e^2_{1,1} \rangle \hspace{0.5cm}  p_{1,1,2} = \langle e^2_{1,4} |e^2_{1,4}\rangle \hspace{0.5cm} p_{2,1,2} = \langle e^2_{1,7} |e^2_{1,7} \rangle 
\end{equation*}
\begin{equation*}
  p_{0,2,0} = \langle e^0_{2,2} |e^0_{2,2} \rangle \hspace{0.5cm}  p_{1,2,0} = \langle e^0_{2,5} |e^0_{2,5}\rangle \hspace{0.5cm}  p_{2,2,0} = \langle e^0_{2,8} |e^0_{2,8}\rangle 
\end{equation*}
\begin{equation*}
  p_{0,2,1} = \langle e^1_{2,2} |e^1_{2,2} \rangle \hspace{0.5cm}  p_{1,2,1} = \langle e^1_{2,5} |e^1_{2,5}\rangle \hspace{0.5cm} p_{2,2,1} = \langle e^1_{2,8} |e^1_{2,8}\rangle  
\end{equation*}
\begin{equation}
  p_{0,2,2} = \langle e^2_{2,2} |e^2_{2,2} \rangle \hspace{0.5cm} p_{1,2,2} = \langle e^2_{2,5} |e^2_{2,5}\rangle \hspace{0.5cm} p_{2,2,2} = \langle e^2_{2,8} |e^2_{2,8}\rangle .
\end{equation}
Given $\rho_{BEC}$, we may easily compute $S(B|EC)= S(BEC)-S(EC)$ now using the above notations. Indeed, it is easy to see that

\begin{equation}
    S(BEC)=S(\rho_{BEC})=H(\frac{1}{3} p_{0,0,0},\frac{1}{3} p_{0,0,1},..., \frac{1}{3} p_{2,2,2}).
\end{equation}

 Next, to compute $S(EC)$, let us introduce the following notations 
\begin{eqnarray}
\sigma_1 &=&|e^0_{0,0}\rangle \langle e^0_{0,0}|+| e^1_{1,4} \rangle \langle  e^1_{1,4}|+ |e^2_{2,8} \rangle \langle  e^2_{2,8}|\nonumber \\
 \sigma_2&=&|e^0_{0,3}\rangle \langle e^0_{0,3}|+|e^0_{0,6}\rangle \langle e^0_{0,6}|+|e^1_{1,1} \rangle \langle e^1_{1,1}|+| e^1_{1,7} \rangle \langle  e^1_{1,7}|+| e^2_{2,2} \rangle \langle  e^2_{2,2}|+| e^2_{2,5} \rangle \langle  e^2_{2,5}|\nonumber \\
 \sigma_3&=&| e^1_{0,0} \rangle \langle  e^1_{0,0}|+|e^2_{0,0} \rangle \langle  e^2_{0,0}|+|e^0_{1,4}\rangle \langle e^0_{1,4}|+| e^2_{1,4} \rangle \langle  e^2_{1,4}|+|e^0_{2,8}\rangle \langle e^0_{2,8}|+| e^1_{2,8} \rangle \langle  e^1_{2,8}|\nonumber \\
 \sigma_4&=&|e^1_{0,3} \rangle\langle e^1_{0,3}|+| e^2_{0,3} \rangle \langle e^2_{0,3}|+| e^1_{0,6} \rangle \langle  e^1_{0,6}|+| e^2_{0,6} \rangle \langle  e^2_{0,6}|+|e^0_{1,1}\rangle \langle e^0_{1,1}|+| e^2_{1,1} \rangle \langle e^2_{1,1}| \nonumber \\
 &&+|e^0_{1,7}\rangle \langle e^0_{1,7}|+| e^2_{1,7} \rangle \langle  e^2_{1,7}|+|e^0_{2,2}\rangle \langle e^0_{2,2}|+| e^1_{2,2} \rangle \langle  e^1_{2,2}|+|e^0_{2,5}\rangle \langle e^0_{2,5}|+| e^1_{2,5} \rangle \langle  e^1_{2,5}|  
\end{eqnarray}

Furthermore, we define $t_j = tr\sigma_j > 0$, for all $j = 1, 2, 3, 4$ with $t_1$ representing the total probability that there is no error between Alice and Bob in both channels (forward and backward), $t_2$ the total probability that there is an error in the forward channel, $t_3$ the total probability that there is an error in the backward channel and $t_4$ is the total probability that there is an error in both channels (forward and backward). We can write the normalized, positive, semi-definite operators now as $\Tilde{\sigma_j}=\sigma_j/ t_j $.

Using this notations, tracing out Bob's state from the density operator $\rho_{BEC}$ in (\ref{rhobec}) yields 
\begin{equation}
     \rho_{EC}  =(\frac{1}{3} t_1 \Tilde{\sigma_1}) \otimes |c,0 \rangle \langle c, 0| + (\frac{1}{3} t_2 \Tilde{\sigma_2}) \otimes |c,1 \rangle \langle c, 1| + (\frac{1}{3} t_3 \Tilde{\sigma_3}) \otimes |w,1 \rangle \langle w, 1| + (\frac{1}{3} t_4 \Tilde{\sigma_4}) \otimes |w,2 \rangle \langle w, 2|.
\end{equation}

Applying Lemma \ref{lemma1} at this point will allow us to write  $S(EC)$ as  
\begin{equation}
\label{sec}
    S(EC)=S(\rho_{EC})=H(\frac{1}{3} t_1,...,\frac{1}{3}t_4)+ \frac{1}{3}\sum^4_{j=1} t_j S(\Tilde{\sigma_j}).
\end{equation}

A lower bound on the key rate equation (\ref{keyrateinf}), requires an upper bound on $S(EC)$, which we can find easily using (\ref{sec}):
\begin{equation}
    S(EC)\leq H\left(\frac{1}{3}t_1,\frac{1}{3}t_2,\frac{1}{3}t_3, \frac{1}{3}t_4\right) +\frac{1}{3}(t_2+t_3+t_4)+\frac{1}{3}t_1 S(\Tilde{\sigma_1}).
\end{equation}

Evidently if the noise of the quantum channel is low, then $p_{i,j,k}$ should be low except for $p_{0,0,0}$, $p_{1,1,1}$ and $p_{2,2,2}$ that should be high.

Next we introduce a new basis $\left\{|a\rangle,|b\rangle,|c\rangle\right\}$ and without loss of generality, we write
\begin{eqnarray}
    |e^0_{0,0}\rangle&=&\chi|a\rangle+\sigma|b\rangle+\rho|c\rangle, \nonumber \\
    |e^1_{1,4}\rangle&=&\sqrt{p_{1,1,1}}|c\rangle,  \\
    |e^2_{2,8}\rangle&=&\sqrt{p_{2,2,2}}|c\rangle, \nonumber \\
\end{eqnarray}
where $\langle a|a\rangle=\langle b|b\rangle=\langle c|c\rangle=1$, $\langle a|b\rangle=\langle b|c\rangle=\langle a|c\rangle=0$, and $\chi,\sigma,\rho \in \mathbb{C}$. This further implies
\begin{equation*}
    |\chi|^2+|\sigma|^2+|\rho|^2=\langle e^0_{0,0}|e^0_{0,0}\rangle=p_{0,0,0}.
\end{equation*}
In this basis, we may write $\Tilde{\sigma_1}$ as
\begin{displaymath} 
\Tilde{\sigma_1}=\frac{1}{p_{0,0,0}+p_{1,1,1}+p_{2,2,2}}\left[\begin{array}{ccc}
|\chi|^2 & \hspace{1 cm} \chi\sigma^{*} & \chi\rho^{*}\\
\sigma\chi^{*} & \hspace{1 cm}  |\sigma|^2 &\sigma\rho^{*}\\
\rho\chi^{*} & \hspace{1 cm} \rho\sigma^{*} & \hspace{1 cm} p_{1,1,1}+p_{2,2,2}+|\rho|^2
\end{array}\right].
\end{displaymath}
 After some algebraic manipulation, the eigenvalues of $\Tilde{\sigma_1}$, denoted $\Tilde{\lambda_0}$, $\Tilde{\lambda_1}$ and $\Tilde{\lambda_2}$ can be found:
\begin{align}
\Tilde{\lambda_0}&= 0, \\
\Tilde{\lambda_1}&=\frac{1}{2}+\frac{\sqrt{4p+p_{0,0,0}^2-2 p_{0,0,0}p_{1,1,1}+p_{1,1,1}^2-2 p_{0,0,0}p_{2,2,2}-2 p_{1,1,1}p_{2,2,2}+p_{2,2,2}^2}}{2(p_{0,0,0}+p_{1,1,1}+p_{2,2,2})}, \label{lambda1tilde} \\
\Tilde{\lambda_2}&=\frac{1}{2}-\frac{\sqrt{4p+p_{0,0,0}^2-2 p_{0,0,0}p_{1,1,1}+p_{1,1,1}^2-2 p_{0,0,0}p_{2,2,2}-2 p_{1,1,1}p_{2,2,2}+p_{2,2,2}^2}}{2(p_{0,0,0}+p_{1,1,1}+p_{2,2,2})},\label{lambda2tilde}\\
\nonumber
\end{align}
where $p=\left(|\langle e^0_{0,0}|e^1_{1,4}\rangle|^2+|\langle e^0_{0,0}|e^2_{2,8}\rangle|^2+|\langle e^1_{1,4}|e^2_{2,8}\rangle|^2\right)$.

\par Incorporating everything together yields the following upper bound on $S(EC)$ 
\begin{equation}
    S(EC)\leq H\left(\frac{1}{3}t_1,\frac{1}{3}t_2,\frac{1}{3}t_3, \frac{1}{3}t_4\right) +\frac{1}{3}(t_2+t_3+t_4)+\frac{1}{3}t_1\left(S(\Tilde{\lambda_1})+S(\Tilde{\lambda_2})\right).
\end{equation}

At first glance, estimating this upper bound based on the expressions of the eigenvalues (\ref{lambda1tilde},\ref{lambda2tilde}) depend on the values $p_{i,j,k}$ but also on the quantity $p$ which cannot be directly observed. However, by using the error rate in the alternative basis ($\mathcal{T}$ or $\mathcal{K}$ depending on the protocol adopted), Alice and Bob may determine bounds on these quantities.

\subsection{Estimating the $\mathcal{T}$ basis noise:}

In order to estimate the channel noise in the basis $\mathcal{T}$ we need to bound the quantity $p$ defined above. For this, we will consider those iterations where Alice choose to encode her qutrit in the  basis $\mathcal{T}$  (\textit{i.e.} she sends the state $|0^\prime \rangle,|1^\prime \rangle$ or $|2^\prime \rangle$), Bob chooses to reflect, and Alice measures in the basis $\mathcal{T}$.

First, we note that the inner product is bounded as follows
\begin{equation*}
    |\langle e^0_{0,0}|e^1_{1,4}\rangle|^2 \geq Re^2(\langle e^0_{0,0}|e^1_{1,4}\rangle),
\end{equation*}
and similarly for the others, 
\begin{equation*}
    |\langle e^0_{0,0}|e^2_{2,8}\rangle|^2\geq Re^2(\langle e^0_{0,0}|e^2_{2,8}\rangle)
\end{equation*}
and
\begin{equation*}
    |\langle e^1_{1,4}|e^2_{2,8}\rangle|^2\geq Re^2(\langle e^1_{1,4}|e^2_{2,8}\rangle).
\end{equation*}

To upper-bound $S(EC)$ requires a lower bound on the quantities $Re^2(\langle e^0_{0,0}|e^1_{1,4}\rangle)+ Re^2(\langle e^0_{0,0}|e^2_{2,8}\rangle)+Re^2(\langle e^1_{1,4}|e^2_{2,8}\rangle)$. Since Bob chooses the operation $R$, the two way quantum channel becomes, essentially, a one way channel with Eve attacking through the unitary operator $V=U_RU_F$. As before, assuming, without loss of generality, that Eve's ancilla is cleared to the zero state $|0\rangle_E$, thus her action on basis states can be described as follows:
\begin{align}
V|0,0\rangle &= U_R(|0,e_0\rangle +|1,e_1\rangle+|2,e_2\rangle) \nonumber \\
&=|0\rangle\otimes\underbrace{(|e^0_{0,0} \rangle + |e^0_{1,1}\rangle+ |e^0_{2,2}\rangle)}_{|f_0\rangle}+|1\rangle\otimes\underbrace{(|e^1_{0,0} \rangle + |e^1_{1,1}\rangle+ |e^1_{2,2}\rangle)}_{|f_1\rangle}+|2\rangle\otimes\underbrace{(|e^2_{0,0} \rangle + |e^2_{1,1}\rangle+ |e^2_{2,2}\rangle)}_{|f_2\rangle}\nonumber \\
&=|0,f_0\rangle +|1,f_1\rangle+|2,f_2\rangle \label{V00};
\end{align}

\begin{align}
V|1,0\rangle &= U_R(|0,e_3\rangle +|1,e_4\rangle+|2,e_5\rangle)\nonumber \\
&=|0\rangle\otimes(|e^0_{0,3} \rangle + |e^0_{1,4}\rangle+ |e^0_{2,5}\rangle)+|1\rangle\otimes(|e^1_{0,3} \rangle + |e^1_{1,4}\rangle+ |e^1_{2,5}\rangle)+|2\rangle\otimes(|e^2_{0,3} \rangle + |e^2_{1,4}\rangle+ |e^2_{2,5}\rangle)\nonumber \\
&=|0,f_3\rangle +|1,f_4\rangle+|2,f_5\rangle \label{V10};
\end{align}

\begin{align}
V|2,0\rangle &= U_R(|0,e_6\rangle +|1,e_7\rangle +|2,e_8\rangle)\nonumber \\
&=|0\rangle\otimes(|e^0_{0,6} \rangle + |e^0_{1,7}\rangle+ |e^0_{2,8}\rangle)+|1\rangle\otimes(|e^1_{0,6} \rangle + |e^1_{1,7}\rangle+ |e^1_{2,8}\rangle)+|2\rangle\otimes(|e^2_{0,6} \rangle + |e^2_{1,7}\rangle+ |e^2_{2,8}\rangle)\nonumber \\
&=|0,f_6\rangle +|1,f_7\rangle+|2,f_8\rangle \label{V20}.
\end{align}

The unitarity of $U_F$ and $ U_R$ allows to obtain
\begin{equation}
    \langle f_{0} |f_{0}\rangle+ \langle f_{1} |f_{1}\rangle+ \langle f_{2} |f_{2}\rangle=\langle f_{3} |f_{3}\rangle+ \langle f_{4} |f_{4}\rangle+ \langle f_{5} |f_{5}\rangle=\langle f_{6} |f_{6}\rangle+ \langle f_{7} |f_{7}\rangle+ \langle f_{8} |f_{8}\rangle=1,
\end{equation}

\begin{equation*}
    \langle f_{0} |f_{3}\rangle+ \langle f_{1} |f_{4}\rangle+ \langle f_{2} |f_{5}\rangle=\langle f_{3} |f_{6}\rangle+ \langle f_{4} |f_{7}\rangle+ \langle f_{5} |f_{8}\rangle= \langle f_{0} |f_{6}\rangle+ \langle f_{1} |f_{7}\rangle+ \langle f_{2} |f_{8}\rangle=0.
\end{equation*}

On the other hand, using its linearity we can describe the action of $V$ on the basis ${\cal{T}}$:
\begin{eqnarray}
V|0^\prime,0\rangle&=& |0^\prime,g_0\rangle +|1^\prime,g_1\rangle+|2^\prime,g_2\rangle,\nonumber \\
V|1^\prime,0\rangle &=&|0^\prime,g_3\rangle +|1^\prime,g_4\rangle+|2^\prime,g_5\rangle,\\
V|2^\prime,0\rangle&=&|0^\prime,g_6\rangle +|1^\prime,g_7\rangle+|2^\prime,g_8\rangle.\nonumber
\end{eqnarray}
Here we have introduced the following states
\begin{equation}
\label{gi}
\begin{split}
|g_0\rangle &=\frac{1}{3}(|f_0\rangle+e^{\frac{2i\pi}{3}}|f_1\rangle+e^{\frac{2i\pi}{3}}|f_2\rangle+e^{\frac{-2i\pi}{3}}|f_3\rangle+|f_4\rangle+|f_5\rangle+e^{\frac{-2i\pi}{3}}|f_6\rangle+|f_7\rangle+|f_8\rangle)\\
|g_1\rangle &=\frac{1}{3}(e^{\frac{2i\pi}{3}}|f_0\rangle+|f_1\rangle+e^{\frac{2i\pi}{3}}|f_2\rangle+|f_3\rangle+e^{\frac{-2i\pi}{3}}|f_4\rangle+|f_5\rangle+|f_6\rangle+e^{\frac{-2i\pi}{3}}|f_7\rangle+|f_8\rangle)\\
|g_2\rangle &=\frac{1}{3}(e^{\frac{2i\pi}{3}}|f_0\rangle+e^{\frac{2i\pi}{3}}|f_1\rangle+|f_2\rangle+|f_3\rangle+|f_4\rangle+e^{\frac{-2i\pi}{3}}|f_5\rangle+|f_6\rangle+|f_7\rangle+e^{\frac{-2i\pi}{3}}|f_8\rangle)\\
|g_3\rangle &=\frac{1}{3}(e^{\frac{-2i\pi}{3}}|f_0\rangle+|f_1\rangle+|f_2\rangle+|f_3\rangle+e^{\frac{2i\pi}{3}}|f_4\rangle+e^{\frac{2i\pi}{3}}|f_5\rangle+e^{\frac{-2i\pi}{3}}|f_6\rangle+|f_7\rangle+|f_8\rangle)\\
|g_4\rangle &=\frac{1}{3}(|f_0\rangle+e^{\frac{-2i\pi}{3}}|f_1\rangle+|f_2\rangle+e^{\frac{2i\pi}{3}}|f_3\rangle+|f_4\rangle+e^{\frac{2i\pi}{3}}|f_5\rangle+|f_6\rangle+e^{\frac{-2i\pi}{3}}|f_7\rangle+|f_8\rangle)\\
|g_5\rangle &=\frac{1}{3}(|f_0\rangle+|f_1\rangle+e^{\frac{-2i\pi}{3}}|f_2\rangle+e^{\frac{2i\pi}{3}}|f_3\rangle+e^{\frac{2i\pi}{3}}|f_4\rangle+|f_5\rangle+|f_6\rangle+|f_7\rangle+e^{\frac{-2i\pi}{3}}|f_8\rangle)\\
|g_6\rangle &=\frac{1}{3}(e^{\frac{-2i\pi}{3}}|f_0\rangle+|f_1\rangle+|f_2\rangle+e^{\frac{-2i\pi}{3}}|f_3\rangle+|f_4\rangle+|f_5\rangle+|f_6\rangle+e^{\frac{2i\pi}{3}}|f_7\rangle+e^{\frac{2i\pi}{3}}|f_8\rangle)\\
|g_7\rangle &=\frac{1}{3}(|f_0\rangle+e^{\frac{-2i\pi}{3}}|f_1\rangle+|f_2\rangle+|f_3\rangle+e^{\frac{-2i\pi}{3}}|f_4\rangle+|f_5\rangle+e^{\frac{2i\pi}{3}}|f_6\rangle+|f_7\rangle+e^{\frac{2i\pi}{3}}|f_8\rangle)\\
|g_8\rangle &=\frac{1}{3}(|f_0\rangle+|f_1\rangle+e^{\frac{-2i\pi}{3}}|f_2\rangle+|f_3\rangle+|f_4\rangle+e^{\frac{-2i\pi}{3}}|f_5\rangle+e^{\frac{2i\pi}{3}}|f_6\rangle+e^{\frac{2i\pi}{3}}|f_7\rangle+|f_8\rangle)
\end{split}
\end{equation}

Consider $\langle g_1|g_1\rangle$ the probability that Alice measures $|1^\prime\rangle$ if she originally sent $|0^\prime\rangle$ ({\textit{i.e.}} the probability $p_{0^\prime1^\prime}$) and $\langle g_2|g_2\rangle$ the probability that Alice measures $|2^\prime\rangle$ if she originally sent $|0^\prime\rangle$ ({\textit{i.e.}} the probability $p_{0^\prime2^\prime}$). $\langle g_{3} |g_{3}\rangle$, $\langle g_{5} |g_{5}\rangle$, $\langle g_{6} |g_{6}\rangle$, and $\langle g_{7} |g_{7}\rangle$ may be defined in a similar manner, namely by using respectively $p_{1^\prime0^\prime}$, $p_{1^\prime2^\prime}$, $p_ {2^\prime0^\prime}$, and $p_ {2^\prime1^\prime}$.  These quantities, which Alice will estimate in the parameter estimation stage, represent the error Eve's attack induces in the $\mathcal{T}$ basis. These expressions are cumbersome, and are displayed in equations (\ref{p01}-\ref{p21}) in appendix \ref{appA}.

Introducing the following notation
\begin{equation}
    \mathcal{S} = \begin{cases} \phantom{-} X_{\phi_1}^2& \text{if } X_{\phi_1} 
\geq 0 \\ \phantom{-} 0 & otherwise \end{cases},
\end{equation}
where
\begin{equation}
    X_{\phi_1}=Re(\langle e^0_{0,0}|e^1_{1,4}\rangle)+Re(\langle e^0_{0,0}|e^2_{2,8}\rangle)+Re(\langle e^1_{1,4}|e^2_{2,8}\rangle),
\end{equation}
using the lower bound, on this last quantity, derived in (\ref{upperbound}) in appendix \ref{appA} and the fact that $p=|\langle e^0_{0,0}|e^1_{1,4}\rangle+\langle e^0_{0,0}|e^2_{2,8}\rangle+\langle e^1_{1,4}|e^2_{2,8}\rangle|^2 \geq \mathcal{S}$, allow us to lower bound $p$ thus upper bound the von Neumann entropy $S(EC)$ and ultimately allow us to find a lower bound on the conditional entropy $S(B|EC)$ in (\ref{strongsubadditivity}).

\subsection{Estimating the $\mathcal{K}$ basis noise:}
In case the protocol is carried out with Alice allowed to choosing her states from set $\Phi_2$, the noise in the $\cal{K}$ basis can be estimated following the same procedure as in the previous subsection. As a matter of fact, the quantity $p=|\langle e^0_{0,0}|e^1_{1,4}\rangle|^2+|\langle e^0_{0,0}|e^2_{2,8}\rangle|^2+|\langle e^1_{1,4}|e^2_{2,8}\rangle|^2$ can be bound by considering those instances where Alice initially sends, then measures in the $\mathcal{K}$ basis while Bob chooses to reflect the qutrit. In this case, Bob's operation is essentially the identity operator, and Eve's action is again the unitary operation $V=U_RU_F$.

Using the linearity of $V$ and the actions of the operators $U_F$ and $U_R$ (\ref{UF},\ref{ur}) we write Eve's effect on the $\mathcal{K}$ basis' states and get equations similar to (\ref{V00}-\ref{V20}) (recall that Eve's ancilla is cleared to the zero state $|0\rangle_E$):
\begin{eqnarray}
V|0^{\prime\prime},0\rangle= |0^{\prime\prime},h_0\rangle +|1^{\prime\prime},h_1\rangle+|2^{\prime\prime},h_2\rangle,\nonumber \\
V|1^{\prime\prime},0\rangle =|0^{\prime\prime},h_3\rangle +|1^{\prime\prime},h_4\rangle+|2^{\prime\prime},h_5\rangle,\\
V|2^{\prime\prime},0\rangle=|0^{\prime\prime},h_6\rangle +|1^{\prime\prime},h_7\rangle+|2^{\prime\prime},h_8\rangle, \nonumber
\end{eqnarray}
where:
\begin{equation}
    \label{hi}
    \begin{split}
        |h_1\rangle &=\frac{1}{3}(|f_0\rangle+e^{\frac{-2i\pi}{3}}|f_1\rangle+e^{\frac{2i\pi}{3}}|f_2\rangle+|f_3\rangle+e^{\frac{-2i\pi}{3}}|f_4\rangle+e^{\frac{2i\pi}{3}}|f_5\rangle+|f_6\rangle+e^{\frac{-2i\pi}{3}}|f_7\rangle+e^{\frac{2i\pi}{3}}|f_8\rangle)\\
        |h_2\rangle &=\frac{1}{3}(|f_0\rangle+e^{\frac{2i\pi}{3}}|f_1\rangle+e^{\frac{-2i\pi}{3}}|f_2\rangle+|f_3\rangle+e^{\frac{2i\pi}{3}}|f_4\rangle+e^{\frac{-2i\pi}{3}}|f_5\rangle+|f_6\rangle+e^{\frac{2i\pi}{3}}|f_7\rangle+e^{\frac{-2i\pi}{3}}|f_8\rangle)\\
        |h_3\rangle &=\frac{1}{3}(|f_0\rangle+|f_1\rangle+|f_2\rangle+e^{\frac{2i\pi}{3}}|f_3\rangle+e^{\frac{2i\pi}{3}}|f_4\rangle+e^{\frac{2i\pi}{3}}|f_5\rangle+e^{\frac{-2i\pi}{3}}|f_6\rangle+e^{\frac{-2i\pi}{3}}|f_7\rangle+e^{\frac{-2i\pi}{3}}|f_8\rangle)\\
        |h_5\rangle &=\frac{1}{3}(|f_0\rangle+e^{\frac{2i\pi}{3}}|f_1\rangle+e^{\frac{-2i\pi}{3}}|f_2\rangle+e^{\frac{2i\pi}{3}}|f_3\rangle+e^{\frac{-2i\pi}{3}}|f_4\rangle+|f_5\rangle+e^{\frac{-2i\pi}{3}}|f_6\rangle+|f_7\rangle+e^{\frac{2i\pi}{3}}|f_8\rangle)\\
        |h_6\rangle &=\frac{1}{3}(|f_0\rangle+|f_1\rangle+|f_2\rangle+e^{\frac{-2i\pi}{3}}|f_3\rangle+e^{\frac{-2i\pi}{3}}|f_4\rangle+e^{\frac{-2i\pi}{3}}|f_5\rangle+e^{\frac{2i\pi}{3}}|f_6\rangle+e^{\frac{2i\pi}{3}}|f_7\rangle+e^{\frac{2i\pi}{3}}|f_8\rangle)\\
        |h_7\rangle &=\frac{1}{3}(|f_0\rangle+e^{\frac{-2i\pi}{3}}|f_1\rangle+e^{\frac{2i\pi}{3}}|f_2\rangle+e^{\frac{-2i\pi}{3}}|f_3\rangle+e^{\frac{2i\pi}{3}}|f_4\rangle+|f_5\rangle+e^{\frac{2i\pi}{3}}|f_6\rangle+|f_7\rangle+e^{\frac{-2i\pi}{3}}|f_8\rangle)\\
    \end{split}
\end{equation}
here we have omitted showing  $|h_0\rangle$,  $|h_4\rangle$ and  $|h_8\rangle$ as they are irrelevant for the calculations afterwards.

Alone the same lines as in the previous subsection we introduce the probabilities $p_{i^{\prime\prime}j^{\prime\prime}}$ for $\left\{i^{\prime\prime}, j^{\prime\prime} \right\}= \left\{0^{\prime\prime}, 1^{\prime\prime}, 2^{\prime\prime} \right\}$ , the probability that Alice measures the returned qutrit in the state $|j^{\prime\prime}\rangle$ when she originally prepared it in the state $|i^{\prime\prime}\rangle$. The expressions of these probabilities are given by equations (\ref{pijprimeprime1}-\ref{pijprimeprime2}) in appendix \ref{appB}.

Introducing notation adapted for this protocol as 
\begin{equation}
    \mathcal{S} = \begin{cases} \phantom{-} X_{\phi_2}^2& \text{if } X_{\phi_2} 
\geq 0 \\ \phantom{-} 0 & otherwise \end{cases},
\end{equation}
with
\begin{equation}
    X_{\phi_2}=Re(\langle e^0_{0,0}|e^1_{1,4}\rangle)+Re(\langle e^0_{0,0}|e^2_{2,8}\rangle)+Re(\langle e^1_{1,4}|e^2_{2,8}\rangle).
\end{equation}

The lower bound on this quantity is shown in equation (\ref{upperbound2}) in appendix \ref{appB} and this allow us to place a lower bound on the conditional entropy $S(B|EC)$ in (\ref{strongsubadditivity}).

\subsection{Final Key Rate Bound}
After parameter estimation, computing $H(B|A)=H(B,A)-H(A)$, is straightforward. Indeed, let $p(b,a)$ be the probability that Bob's trit is $b$ while Alice's is $a$. These probabilities are given by
\begin{equation}
\begin{split}
   p(0,0)&=\frac{1}{3}(p_{000}+p_{100}+p_{200}) \hspace{0.5cm} p(0,1)=\frac{2}{3}(p_{001}+p_{101}+p_{201}) \hspace{0.5 cm} p(0,2)=\frac{2}{3}(p_{002}+p_{102}+p_{202})\\
    p(1,0)&=\frac{2}{3}(p_{010}+p_{110}+p_{210})\hspace{0.5 cm} p(1,1)=\frac{1}{3}(p_{011}+p_{111}+p_{211})\hspace{0.5 cm} p(1,2)=\frac{2}{3}(p_{012}+p_{112}+p_{212})\\
    p(2,0)&=\frac{2}{3}(p_{020}+p_{120}+p_{220})\hspace{0.5 cm}p(2,1)=\frac{2}{3}(p_{021}+p_{121}+p_{221})\hspace{0.5 cm}p(2,2)=\frac{1}{3}(p_{022}+p_{122}+p_{222})
    \end{split}
\end{equation}
Also,
let $p_A(0)$ be the probability that Alice's trit is zero, $p_A(1)$ and $p_A(2)$  be the probability that it is one, and two respectively. Then we have
\begin{equation}
\begin{split}
    p_A(0)&=\frac{1}{3}p_{000}+\frac{2}{3}p_{010}+\frac{2}{3}p_{020}+\frac{2}{3}p_{110}+\frac{1}{3}p_{100}+\frac{2}{3}p_{120}+\frac{1}{3}p_{200}+\frac{2}{3}p_{210}+\frac{2}{3}p_{220}\\
    p_A(1)&=\frac{2}{3}p_{001}+\frac{1}{3}p_{011}+\frac{2}{3}p_{021}+\frac{2}{3}p_{101}+\frac{1}{3}p_{111}+\frac{2}{3}p_{121}+\frac{2}{3}p_{201}+\frac{1}{3}p_{211}+\frac{2}{3}p_{221}\\
    p_A(2)&=\frac{2}{3}p_{002}+\frac{2}{3}p_{012}+\frac{1}{3}p_{022}+\frac{2}{3}p_{102}+\frac{2}{3}p_{112}+\frac{1}{3}p_{122}+\frac{2}{3}p_{202}+\frac{2}{3}p_{212}+\frac{1}{3}p_{222}
\end{split}
\end{equation}

Putting everything together, the key rate bound is found to be
\begin{equation}
\label{keyratebound}
\begin{split}
    r \geq& H(\frac{1}{3} p_{0,0,0},\frac{1}{3} p_{0,0,1},..., \frac{1}{3} p_{2,2,2})- H\left(\frac{1}{3}t_1,\frac{1}{3}t_2,\frac{1}{3}t_3, \frac{1}{3}t_4\right) -\frac{1}{3}(t_2+t_3+t_4)\\ &-\frac{1}{3}t_1\left(H(\Tilde{\lambda_1})+H(\Tilde{\lambda_2})\right)+H(p_A(0),p_A(1),p_A(2))\\ 
    & - H(p(0,0),p(0,1),p(0,2),p(1,0),p(1,1),p(1,2),p(2,0),p(2,1),p(2,2)). 
\end{split}
\end{equation}
 
In this equation, the $\Tilde{\lambda}$'s from equations (\ref{lambda1tilde}) and (\ref{lambda2tilde})  are functions of $\mathcal{S}$, depending only on the parameters that Alice and Bob will estimate.

\subsection{General Attacks}

In the above, we considered only one attack termed as a collective attack. However, the protocol considered in this paper may be made permutation invariant, by permuting the raw key using a random permutation chosen publicly. In this context, the security of a (S)QKD scheme against general attacks follows directly from its security against collective attacks (see \cite{PhysRevLett.102.020504,R1} ). Thus, in the asymptotic scenario, our key rate bound will still be the same.

\subsection{Results and discussion}

\par To evaluate the key-rate (\ref{keyratebound}), we consider two common forms of channel attacks (noise), \textit{independent channel} and \textit{dependent channel}. For the former, we assume that $Q_{indep} = 2Q(2-3Q) $ (\textit{i.e.} the $\mathcal{T}$ or  $\mathcal{K}$ basis noise observed when the qutrit travels through both channels whenever Bob chooses Reflect). In the later, the observed $\mathcal{T}$ or $\mathcal{K}$ basis noise is simply $Q$ (\textit{i.e.} $Q_{dep} = Q$ ).

\begin{figure}[ht]
 
  \checkoddpage
 \edef\side{\ifoddpage l\else r\fi}%
  \makebox[\textwidth][\side]{%
    \begin{minipage}[t]{0.47\textwidth}
      \centering
      \includegraphics[width=\textwidth]{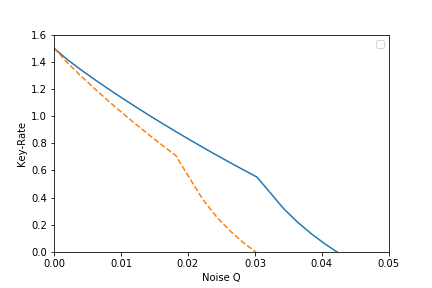}
     \caption{Lower bound on the key rate for  the $\Phi_2$-SQKD protocol as a function of the noise Q. The dashed line represents the independent channel, whereas the solid line shows the dependent channel.}
     \label{rvsQphi2}
    \end{minipage}%
  \hfill
    \begin{minipage}[t]{0.47\textwidth}
     \centering
      \includegraphics[width=\textwidth]{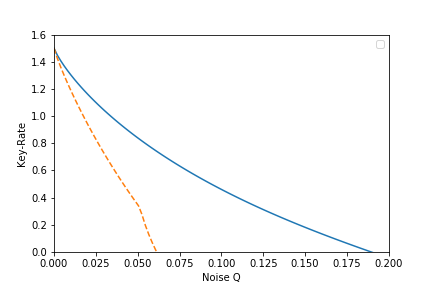}
      \caption{The key rate for the $\Phi_1$-SQKD protocol as a function of the noise Q. For the dashed line, We consider $Q_T = 2Q(2-3Q)$ (independent channel), whereas for the solid line, we consider $Q_T = Q$ (dependent channel)}
      \label{rvsQphi1}
    \end{minipage}%
  }%
\end{figure}

The behavior of the lower bound of the key rate $r$ as a function of the parameter Q for the 3-dimensional SQKD protocol, when  Alice is choosing her qutrit states from the set $\Phi_2$, is shown in Figure \ref{rvsQphi2}. In the first case (the independent channel) the key rate is positive for all Q$\leq 3\%$. In the second case (the dependent channel), the key rate remains positive for all Q$\leq 4.2\%$. 

The same behavior when Alice chooses her states from the set $\Phi_1$, is shown in Figure \ref{rvsQphi1}. In this graph, we observe that the maximal noise tolerance for a dependent channel is $19.1\%$ and it drops to $6.1\%$ for an independent channel.

Interestingly, the $\Phi_1$-SQKD protocol gives higher key rate than the $\Phi_2$-SQKD protocol. Moreover, when $Q_T = Q$, the noise tolerance of our protocol is higher than the $5.34\%$ allowed by the 2-dimensional SQKD protocol \cite{2U}. It is also higher than the noise tolerance of the QKD protocol which can tolerate up to $15\%$ noise \cite{1F}. This shows that the high dimensional advantage, known for fully quantum protocols, carries also to the semi-quantum models.

\section {Conclusions}

It has been identified that in Quantum key distribution, higher dimension not only enhances the key rate but it is also known to be more resilient to errors. Motivated by this we try to understand whether this is the case for Semi quantum key distribution protocols as well. In this direction, we presented a security analysis of a 3-dimensional semi quantum protocol based on two mutually unbiased bases.

In this paper, we have provided a proof of unconditional security for a semi-quantum key distribution protocol based on 3-dimensional quantum states using the technique of conditioning on the additional random variable. We have proved the security of the protocol against collective attacks and argued that it is valid for more general attacks. Moreover, we have derived a lower bound on the key rate, in the asymptotic scenario, as function only of the quantum channel’s noise (a parameter that may be estimated by Alice and Bob). In order to evaluate its lower bound we have considered the most common forms of channels; namely the independent  channel and dependent  channel. Our results show that 3d-SQKD undoubtedly provides better security than the original qubit SQKD. Thus, similar to the fully quantum key distribution protocol, moving to higher dimensions is advantageous in the case of semi-quantum key distribution protocol. It will be interesting to study how the protocol carries out if we move to even higher dimensions and considering a $d$-dimensional version of the protocol.

\section*{Acknowledgements}

This work has been supported by the National Center for Scientific and Technical Research (CNRST). 

\noindent The authors would like to thank W. O. Krawec for fruitful discussions and help on his original work.

\appendix
\appendixpage
\renewcommand{\theequation}{\thesection.\arabic{equation}}
\section{Bounding the entropy in the $\mathcal{T}$ basis}
\label{appA}
From equation (\ref{gi}) one can find the expressions of the quantities $p_{i^\prime j^\prime}$:
\begin{equation}
\label{p01}
\begin{split}
p_{0^\prime1^\prime}=&\langle g_{1}|g_{1}\rangle\\ =& \frac{1}{3}+\frac{1}{9}Re(\langle f_{3}|f_{1}\rangle+\langle f_{5}| f_{1}\rangle+\langle f_{6}|f_{1}\rangle+\langle f_{8}|f_{1}\rangle+\langle f_{1}|f_{3}\rangle+\langle f_{5}| f_{3}\rangle+\langle f_{8}|f_{3}\rangle+\langle f_{1}|f_{5}\rangle\\
& \quad +\langle f_{3} |f_{5}\rangle+\langle f_{6}|f_{5}\rangle+\langle f_{1} |f_{6}\rangle+\langle f_{5} |f_{6}\rangle+\langle f_{8} |f_{6}\rangle+\langle f_{1} |f_{8}\rangle+\langle f_{3} |f_{8}\rangle+\langle f_{6} |f_{8}\rangle+\langle f_{2} |f_{0}\rangle+\langle f_{0} |f_{2}\rangle)\\
& \quad +\frac{1}{9}e^{\frac{-2i\pi}{3}}Re(\langle f_{0}| f_{1}\rangle+\langle f_{2} |f_{1}\rangle+\langle f_{2} |f_{3}\rangle+\langle f_{0} |f_{5}\rangle+\langle f_{2}| f_{6}\rangle+\langle f_{0} |f_{8}\rangle+\langle f_{4} |f_{0}\rangle+\langle f_{7} |f_{0}\rangle \\
& \quad +\langle f_{4} f_{2}\rangle+\langle f_{7} |f_{2}\rangle+\langle f_{3} |f_{4}\rangle+\langle f_{5} |f_{4}\rangle+\langle f_{6}| f_{4}\rangle+\langle f_{8} |f_{4}\rangle+\langle f_{3} |f_{7}\rangle+\langle f_{5} |f_{7}\rangle+\langle f_{6} |f_{7}\rangle+\langle f_{8} |f_{7}\rangle)\\
& \quad +\frac{1}{9}e^{\frac{2i\pi}{3}}Re(\langle f_{4} |f_{3}\rangle+\langle f_{7} |f_{3}\rangle+\langle f_{4} |f_{5}\rangle+\langle f_{7} |f_{5}\rangle+\langle f_{4} |f_{6}\rangle+\langle f_{7} |f_{6}\rangle+\langle f_{4} |f_{8}\rangle+\langle f_{7} |f_{8}\rangle+\langle f_{1} |f_{0}\rangle \\
& \quad +\langle f_{5} |f_{0}\rangle+\langle f_{8} |f_{0}\rangle+\langle f_{1} |f_{2}\rangle+\langle f_{3} |f_{2}\rangle+\langle f_{6} |f_{2}\rangle+\langle f_{0} |f_{4}\rangle+\langle f_{2} |f_{4}\rangle+\langle f_{0} |f_{7}\rangle+\langle f_{2} |f_{7}\rangle)
\end{split}
\end{equation}

\begin{equation}
\begin{split}
p_{0^\prime2^\prime}=&\langle g_{2} |g_{2}\rangle\\ =&\frac{1}{3}+\frac{1}{9}Re(\langle f_{1} |f_{0}\rangle+\langle f_{0}| f_{1}\rangle+\langle f_{3} |f_{2}\rangle+\langle f_{4} |f_{2}\rangle+\langle f_{6} |f_{2}\rangle+\langle f_{7}| f_{2}\rangle+\langle f_{2} |f_{3}\rangle+\langle f_{4} |f_{3}\rangle\\
& \quad +\langle f_{7} |f_{3}\rangle+\langle f_{2}|f_{4}\rangle+\langle f_{3} |f_{4}\rangle+\langle f_{6} |f_{4}\rangle+\langle f_{2} |f_{6}\rangle+\langle f_{4} |f_{6}\rangle+\langle f_{7} |f_{6}\rangle+
\langle f_{2} |f_{7}\rangle+\langle f_{3} |f_{7}\rangle+\langle f_{6} |f_{7}\rangle)\\
& \quad +\frac{1}{9}e^{\frac{2i\pi}{3}}Re(\langle f_{2} |f_{0}\rangle+\langle f_{4} |f_{0}\rangle+\langle f_{7} |f_{0}\rangle+\langle f_{2} |f_{1}\rangle+\langle f_{3}| f_{1}\rangle+\langle f_{6} |f_{1}\rangle+\langle f_{5} |f_{3}\rangle+\langle f_{8} |f_{3}\rangle \\
& \quad +\langle f_{5} f_{4}\rangle+\langle f_{8} |f_{4}\rangle+\langle f_{0} |f_{5}\rangle+\langle f_{1} |f_{5}\rangle+\langle f_{5} |f_{6}\rangle+\langle f_{8} |f_{6}\rangle+\langle f_{5} |f_{7}\rangle+\langle f_{8} |f_{7}\rangle+\langle f_{0}| f_{8}\rangle+\langle f_{1} |f_{8}\rangle)\\
& \quad  +\frac{1}{9}e^{\frac{-2i\pi}{3}}Re(\langle f_{5} |f_{0}\rangle+\langle f_{8} |f_{0}\rangle+\langle f_{5} |f_{1}\rangle+\langle f_{8} |f_{1}\rangle+\langle f_{0} |f_{2}\rangle+\langle f_{1} |f_{2}\rangle+\langle f_{1} |f_{3}\rangle+\langle f_{0} |f_{4}\rangle+\langle f_{3} |f_{5}\rangle \\
& \quad +\langle f_{4} |f_{5}\rangle+\langle f_{6} |f_{5}\rangle+\langle f_{7} |f_{5}\rangle+\langle f_{1} |f_{6}\rangle+\langle f_{0} |f_{7}\rangle+\langle f_{3} |f_{8}\rangle+\langle f_{4} |f_{8}\rangle+\langle f_{6} |f_{8}\rangle+\langle f_{7} |f_{8}\rangle)
\end{split}
\end{equation}

\begin{equation}
\begin{split}
p_{1^\prime0^\prime}=&\langle g_{3} |g_{3}\rangle\\ =&\frac{1}{3}+\frac{1}{9}Re(\langle f_{2} |f_{1}\rangle+\langle f_{3} |f_{1}\rangle+\langle f_{8} |f_{1}\rangle+\langle f_{1} |f_{2}\rangle+\langle f_{3} |f_{2}\rangle+\langle f_{7} |f_{2}\rangle+\langle f_{1} |f_{3}\rangle+\langle f_{2} |f_{3}\rangle\\
& \quad +\langle f_{7} |f_{3}\rangle+\langle f_{8}|f_{3}\rangle+\langle f_{5} |f_{4}\rangle+\langle f_{4} |f_{5}\rangle+\langle f_{2} |f_{7}\rangle+\langle f_{3} |f_{7}\rangle+\langle f_{8} |f_{7}\rangle+\langle f_{1} |f_{8}\rangle+\langle f_{3} |f_{8}\rangle+\langle f_{7} |f_{8}\rangle)\\
& \quad +\frac{1}{9}e^{\frac{-2i\pi}{3}}Re(\langle f_{1} |f_{0}\rangle+\langle f_{2} |f_{0}\rangle+\langle f_{7} |f_{0}\rangle+\langle f_{8} |f_{0}\rangle+\langle f_{5}| f_{1}\rangle+\langle f_{4} |f_{2}\rangle+\langle f_{4} |f_{3}\rangle+\langle f_{5} |f_{3}\rangle \\
& \quad +\langle f_{0} |f_{4}\rangle+\langle f_{6} |f_{4}\rangle+\langle f_{0} |f_{5}\rangle+\langle f_{6} |f_{5}\rangle+\langle f_{1} |f_{6}\rangle+\langle f_{2} |f_{6}\rangle+\langle f_{7} |f_{6}\rangle+\langle f_{8} |f_{6}\rangle+\langle f_{5} |f_{7}\rangle+\langle f_{4} |f_{8}\rangle)\\
& \quad +\frac{1}{9}e^{\frac{2i\pi}{3}}Re(\langle f_{4} |f_{0}\rangle+\langle f_{5} |f_{0}\rangle+\langle f_{0} |f_{1}\rangle+\langle f_{6} |f_{1}\rangle+\langle f_{0} |f_{2}\rangle+\langle f_{6} |f_{2}\rangle+\langle f_{2} |f_{4}\rangle+\langle f_{3} |f_{4}\rangle+\langle f_{8} |f_{4}\rangle \\
& \quad +\langle f_{1} |f_{5}\rangle+\langle f_{3} |f_{5}\rangle+\langle f_{7} |f_{5}\rangle+\langle f_{4} |f_{6}\rangle+\langle f_{5} |f_{6}\rangle+\langle f_{0} |f_{7}\rangle+\langle f_{6} |f_{7}\rangle+\langle f_{0} |f_{8}\rangle+\langle f_{6} |f_{8}\rangle)
\end{split}
\end{equation}

\begin{equation}
\begin{split}
p_{1^\prime2^\prime} =& \langle g_{5}|g_{5}\rangle\\ =&\frac{1}{3}+\frac{1}{9}Re(\langle f_{1}|f_ {0}\rangle+\langle f_{5}|f_{0}\rangle +\langle f_{7}|f_{0} \rangle+ \langle f_{0}|f_{1}\rangle+\langle f_{5}|f_{1} \rangle+\langle f_{6}|f_{1} \rangle+\langle f_{4}|f_{3}\rangle+\langle f_{3}|f_{4} \rangle+\langle f_{0}|f_{5}\rangle+\langle f_{1}|f_{5} \rangle\\
& \quad +\langle f_{6}|f_{5}\rangle+\langle f_{7}|f_{5} \rangle+\langle f_{1}|f_{6}\rangle+\langle f_{5}|f_{6}\rangle+\langle f_{7}|f_{6}\rangle+\langle f_{0}|f_{7}\rangle+\langle f_{5}|f_{7}\rangle+\langle f_{6}|f_{7} \rangle) \\
& \quad +\frac {1}{9}e^{\frac{-2 i\pi}{3}}Re(\langle f_{4}|f_{0}\rangle+\langle f_{3}|f_{1}\rangle+\langle f_{0}|f_{2} \rangle + \langle f_{1}|f_{2}\rangle+\langle f_{6}|f_{2}\rangle +\langle f_{7}|f_{2} \rangle+\langle f_{2} |f_{3} \rangle+\langle f_{8}|f_{3} \rangle \\
& \quad + \langle f_{2}|f_{4} \rangle + \langle f_{8}|f_{4} \rangle + \langle f_{3}|f_{5} \rangle + \langle f_{4}|f_{5} \rangle + \langle f_{4}|f_{6} \rangle + \langle f_{3}|f_{7} \rangle + \langle f_{0}|f_{8} \rangle + \langle f_{1}|f_{8} \rangle + \langle f_ {6}|f_{8} \rangle +\langle f_{7}|f_{8} \rangle) \\
& \quad + \frac {1} {9} e^{\frac {2 i\pi} {3}}Re(\langle f_ {2} |f_{0} \rangle + \langle f_{8}|f_{0} \rangle + \langle f_{2}|f_{1} \rangle + \langle f_{8}|f_{1} \rangle + \langle f_{3}|f_{2} \rangle + \langle f_{4}|f_{2} \rangle + \langle f_{1}|f_{3} \rangle + \langle f_{5}|f_{3} \rangle + \langle f_{7}|f_{3}\rangle \\
& \quad +\langle f_{0}|f_{4}\rangle + \langle f_{5}|f_{4} \rangle+\langle f_{6}|f_{4} \rangle + \langle f_{2}|f_{6} \rangle + \langle f_{8}|f_{6} \rangle + \langle f_{2}|f_{7} \rangle + \langle f_{8}|f_{7} \rangle + \langle f_{3}|f_{8} \rangle + \langle f_{4}|f_{8}\rangle)
\end{split}
\end{equation}

\begin{equation}
\begin{split}
p_ {2^\prime0^\prime} =& \langle g_{6}|g_{6}\rangle\\ =&\frac{1}{3}+\frac{1}{9}Re(\langle f_{2}|f_{1}\rangle+\langle f_{5}|f_{1}\rangle+\langle f_{6}|f_{1} \rangle+\langle f_{1}|f_{2}\rangle+\langle f_{4}|f_{2}\rangle+\langle f_{6}|f_{2}\rangle+\langle f_{2}|f_{4}\rangle+\langle f_{5}|f_{4}\rangle \\
& \quad +\langle f_{6}|f_{4} \rangle+\langle f_{1}|f_{5} \rangle+\langle f_{4}|f_{5}\rangle+\langle f_{6}|f_{5}\rangle+\langle f_{1}|f_{6}\rangle+\langle f_{2}|f_{6}\rangle+\langle f_{4}|f_{6}\rangle+\langle f_{5}|f_{6}\rangle+\langle f_{8}| f_{7}\rangle+\langle f_{7}|f_{8}\rangle)\\
& \quad + \frac {1}{9}e^{\frac {-2 i\pi}{3}}Re(\langle f_{1}|f_{0}\rangle+\langle f_{2}|f_{0}\rangle+\langle f_{4}|f_{0}\rangle+\langle f_{5}|f_{0}\rangle+\langle f_{8}|f_{1}\rangle+\langle f_{7}|f_{2}\rangle+\langle f_{1}|f_{3}\rangle+\langle f_{2}|f_{3}\rangle \\
& \quad +\langle f_{4}|f_{3}\rangle+\langle f_{5}|f_{3}\rangle+\langle f_{8}|f_{4}\rangle+\langle f_{7}|f_{5}\rangle+\langle f_{7}|f_{6}\rangle+\langle f_{8}|f_{6}\rangle+\langle f_{0}|f_{7}\rangle+\langle f_{3}|f_{7}\rangle+\langle f_{0}|f_{8}\rangle+\langle f_{3}|f_{8}\rangle) \\
& \quad + \frac {1} {9} e^{\frac {2 i\pi}{3}}Re(\langle f_{7}|f_{0}\rangle+\langle f_{8}|f_{0}\rangle+\langle f_{0}|f_{1}\rangle+\langle f_{3}|f_{1}\rangle+\langle f_{0}|f_{2}\rangle+\langle f_{3}|f_{2}\rangle+\langle f_{7}|f_{3}\rangle+\langle f_{8}|f_{3}\rangle+\langle f_{0}|f_{4}\rangle+\langle f_{3}|f_{4}\rangle\\
& \quad +\langle f_{0}|f_{5}\rangle+\langle f_{3}|f_{5}\rangle+\langle f_{0}|f_{6}\rangle+\langle f_{1}|f_{7}\rangle+\langle f_{2}|f_{7}\rangle+\langle f_{5}|f_{7}\rangle+\langle f_{6}|f_{7}\rangle+\langle f_{1}|f_{8}\rangle+\langle f_{2}|f_{8}\rangle+\langle f_{4}|f_{8}\rangle+\langle f_{6}|f_{8}\rangle)
\end{split}
\end{equation}

\begin{equation}
\begin{split}
\label{p21}
p_ {2^\prime1^\prime} =& \langle g_ {7}|g_ {7} \rangle\\ =&\frac {1}{3}+\frac{1}{9}Re(\langle f_{2}|f_{0} \rangle+\langle f_{5}|f_{0} \rangle + \langle f_{7}|f_{0} \rangle + \langle f_{0}|f_{2} \rangle+\langle f_{3}|f_{2}\rangle+\langle f_{7}|f_{2} \rangle + \langle f_{0}|f_{3} \rangle + \langle f_ {2} | f_ {3} \rangle+\langle f_{5}|f_{3} \rangle + \langle f_{7}|f_{3} \rangle \\
& \quad + \langle f_{1}|f_{4} \rangle + \langle f_{0}|f_{5} \rangle + \langle f_{2}|f_{5} \rangle + \langle f_{3}|f_{5} \rangle + \langle f_{7}|f_{5} \rangle+ \langle f_{8}|f_{6} \rangle+ \langle f_{0}|f_{7} \rangle + \langle f_{2}|f_{7} \rangle + \langle f_{3}|f_{7} \rangle + \langle f_{5}|f_{7} \rangle + \langle f_ {6} |  f_ {8} \rangle) \\
& \quad + \frac {1} {9} e^{\frac {-2 i\pi} {3}}Re(\langle f_{8}|f_{0} \rangle + \langle f_{0}|f_{1} \rangle+\langle f_{2}|f_{1} \rangle + \langle f_{3}|f_{1} \rangle+\langle f_{5}|f_{1} \rangle+ \langle f_{6}|f_{2} \rangle + \langle f_{8}|f_{3}\rangle + \langle f_{0}|f_{4} \rangle \\
& \quad + \langle f_{2}|f_{4} \rangle +\langle f_{3}|f_{4} \rangle + \langle f_{5}|f_{4} \rangle + \langle f_{6}|f_{5} \rangle+ \langle f_{1}|f_{6} \rangle + \langle f_{4}|f_{6} \rangle + \langle f_{6}|f_{7}\rangle + \langle f_{8}|f_{7} \rangle + \langle f_ {1}| f_ {8} \rangle + \langle f_ {4}| f_ {8} \rangle) \\
& \quad + \frac {1} {9} e^{\frac {2 i\pi} {3}}Re(\langle f_ {1} | f_ {0} \rangle + \langle f_ {4} | f_ {0} \rangle + \langle f_ {6} | f_ {1} \rangle + \langle f_ {8} | f_ {1} \rangle + \langle f_ {1} | f_ {2} \rangle + \langle f_ {4} | f_ {2} \rangle + \langle f_ {1} | f_ {3} \rangle + \langle f_ {4} | f_ {3} \rangle + \langle f_ {6} | f_ {4} \rangle \\ 
& \quad + \langle f_ {8} |f_ {4} \rangle + \langle f_ {1} | f_ {5} \rangle + \langle f_ {4} | f_ {5} \rangle + \langle f_ {2} | f_ {6} \rangle + \langle f_ {5} | f_ {6} \rangle + \langle f_ {7} | f_ {6} \rangle + \langle f_ {0} | f_ {8} \rangle + \langle f_ {3} | f_ {8} \rangle + \langle f_ {7} | f_ {8} \rangle)
\end{split}
\end{equation}

From this, and after some algebraic manipulation, it follows from Cauchy-Schwarz’s inequality that the quantity $X_{\phi_1}=Re(\langle e^0_{0,0}|e^1_{1,4}\rangle)+Re(\langle e^0_{0,0}|e^2_{2,8}\rangle)+Re(\langle e^1_{1,4}|e^2_{2,8}\rangle)$ can bounded as follows

\begin{equation}
\label{upperbound}
    \begin{split}
    X_{\phi_1}\geq 3&-\frac{3}{2}(p_{0^\prime1^\prime}+p_{0^\prime2^\prime}+p_{1^\prime0^\prime}+p_{1^\prime2^\prime}+p_{2^\prime0^\prime}+p_{2^\prime1^\prime})  \\
&+\frac{1}{2}(\sqrt{p_{001}p_{102}}+\sqrt{p_{011}p_{102}}+\sqrt {p_{021}p_{102}}+\sqrt{p_{001}p_{112}}+\sqrt{p_{011}p_{112}} \\
&+\sqrt {p_{021} p_{112}}+\sqrt{p_{001} p_{122}}+\sqrt{p_{011} p_{122}}+\sqrt {p_{021} p_{122}}+\sqrt {p_{001} p_{200}} \\
&+\sqrt{p_{011} p_{200}}+\sqrt {p_{021} p_{200}}+\sqrt{p_{001} p_{210}}+\sqrt {p_{011} p_{210}}+\sqrt {p_{021} p_{210}} \\
&+\sqrt{p_{001} p_{220}}+\sqrt {p_{011} p_{220}}+\sqrt {p_{021} p_{220}}+\sqrt{p_{002} p_{100}}+\sqrt {p_{012} p_{100}} \\
&+\sqrt {p_{022} p_{100}}+\sqrt{p_{002} p_{110}}+\sqrt {p_{012} p_{110}}+\sqrt {p_{022} p_{110}}+\sqrt{p_{002} p_{120}} \\
&+\sqrt {p_{012} p_{120}}+\sqrt {p_{022} p_{120}}+\sqrt {p_{002} p_{201}}+\sqrt {p_{012} p_{201}}+\sqrt {p_{022} p_{201}} \\
&+\sqrt{p_{002} p_{211}}+\sqrt {p_{012} p_{211}}+\sqrt {p_{022} p_{211}}+\sqrt{p_{002} p_{221}}+\sqrt {p_{012} p_{221}} \\
&+\sqrt {p_{022} p_{221}}+\sqrt{p_{100} p_{201}}+\sqrt {p_{110} p_{201}}+\sqrt {p_{120} p_{201}}+\sqrt{p_{100} p_{211}} \\
&+\sqrt {p_{110} p_{211}}+\sqrt {p_{120} p_{211}}+\sqrt{p_{100} p_{221}}+\sqrt {p_{110} p_{221}}+\sqrt {p_{120} p_{221}} \\
&+\sqrt{p_{102} p_{200}}+\sqrt {p_{112} p_{200}}+\sqrt{p_{122} p_{200}}+\sqrt {p_{102} p_{210}}+\sqrt{p_{112} p_{210}} \\
&+\sqrt {p_{122} p_{210}}+\sqrt{p_{102} p_{220}}+\sqrt {p_{112} p_{220}}+\sqrt {p_{122} p_{220}} ) \\
&-(\sqrt {p_{000} p_{101}}+\sqrt {p_{010} p_{101}}+\sqrt {p_{020} p_{101}}+\sqrt {p_{010} p_{111}}+\sqrt {p_{020} p_{111}} \\
&+\sqrt {p_{000} p_{121}}+\sqrt {p_{010} p_{121}}+\sqrt {p_{020} p_{121}}+\sqrt {p_{000} p_{202}}+\sqrt {p_{010} p_{202}} \\
&+\sqrt {p_{020} p_{202}}+\sqrt {p_{000} p_{212}}+\sqrt {p_{010} p_{212}} +\sqrt {p_{020} p_{212}} + \sqrt {p_{010}p_{222}} \\
&+\sqrt {p_{020} p_{222}}+\sqrt {p_{101} p_{202}}+\sqrt {p_{111} p_{202}} +\sqrt {p_{121} p_{202}}+\sqrt {p_{101} p_{212}}  \\
&+\sqrt {p_{111} p_{212}} + \sqrt {p_{121} p_{212}} +\sqrt {p_{101} p_{222}} + \sqrt {p_{121} p_{222}}).
\end{split}
\end{equation}

\section{Bounding the entropy in the $\mathcal{K}$ basis}
\label{appB}

From equation (\ref{hi}) one can find the expressions of the quantities $p_{i^{\prime\prime} j^{\prime\prime}}$:

\begin{equation}
\label{pijprimeprime1}
    \begin{split}
    p_{0^{\prime\prime}1^{\prime\prime}}=&\langle h_{1}|h_{1}\rangle\\ 
    =&\frac{1}{3} +\frac{1}{9}e^{\frac{2i\pi}{3}}Re(\langle f_{1}| f_{0}\rangle+\langle f_{4} |f_{0}\rangle+\langle f_{7} |f_{0}\rangle+\langle f_{2} |f_{1}\rangle+\langle f_{5}| f_{1}\rangle+\langle f_{8} |f_{1}\rangle+\langle f_{0} |f_{2}\rangle+\langle f_{3} |f_{2}\rangle \\
    & \quad +\langle f_{6} f_{2}\rangle+\langle f_{1} |f_{3}\rangle+\langle f_{4} |f_{3}\rangle+\langle f_{7} |f_{3}\rangle+\langle f_{2}| f_{4}\rangle+\langle f_{5} |f_{4}\rangle+\langle f_{8} |f_{4}\rangle+\langle f_{0} |f_{5}\rangle+\langle f_{3} |f_{5}\rangle+\langle f_{6} |f_{5}\rangle\\
    & \quad +\langle f_{1} |f_{6}\rangle+\langle f_{4} |f_{6}\rangle+\langle f_{7} |f_{6}\rangle+\langle f_{2} |f_{7}\rangle+\langle f_{5} |f_{7}\rangle+\langle f_{8} |f_{7}\rangle+\langle f_{0} |f_{8}\rangle+\langle f_{3} |f_{8}\rangle+\langle f_{6} |f_{8}\rangle)\\
    & \quad  +\frac{1}{9}e^{\frac{-2i\pi}{3}}Re(\langle f_{2} |f_{0}\rangle+\langle f_{5} |f_{0}\rangle+\langle f_{8} |f_{0}\rangle+\langle f_{0} |f_{1}\rangle+\langle f_{3} |f_{1}\rangle+\langle f_{6} |f_{1}\rangle+\langle f_{1} |f_{2}\rangle+\langle f_{4} |f_{2}\rangle+\langle f_{7} |f_{2}\rangle \\
    & \quad +\langle f_{2} |f_{3}\rangle+\langle f_{5} |f_{3}\rangle+\langle f_{8} |f_{3}\rangle+\langle f_{0} |f_{4}\rangle+\langle f_{3} |f_{4}\rangle+\langle f_{6} |f_{4}\rangle+\langle f_{1} |f_{5}\rangle+\langle f_{4} |f_{5}\rangle+\langle f_{7} |f_{5}\rangle\\
    & \quad +\langle f_{2} |f_{6}\rangle+\langle f_{5} |f_{6}\rangle+\langle f_{8} |f_{6}\rangle+\langle f_{0} |f_{7}\rangle+\langle f_{3} |f_{7}\rangle+\langle f_{6} |f_{7}\rangle+\langle f_{1} |f_{8}\rangle+\langle f_{4} |f_{8}\rangle+\langle f_{7} |f_{8}\rangle)    
    \end{split}
\end{equation}

\begin{equation}
    \begin{split}
    p_{0^{\prime\prime}2^{\prime\prime}}=&\langle h_{2}|h_{2}\rangle\\ =&\frac{1}{3}+\frac{1}{9}e^{\frac{2i\pi}{3}}Re(\langle f_{2} |f_{0}\rangle+\langle f_{5} |f_{0}\rangle+\langle f_{8} |f_{0}\rangle+\langle f_{0} |f_{1}\rangle+\langle f_{3} |f_{1}\rangle+\langle f_{6} |f_{1}\rangle+\langle f_{1} |f_{2}\rangle+\langle f_{4} |f_{2}\rangle+\langle f_{7} |f_{2}\rangle \\
& \quad +\langle f_{2} |f_{3}\rangle+\langle f_{5} |f_{3}\rangle+\langle f_{8} |f_{3}\rangle+\langle f_{0} |f_{4}\rangle+\langle f_{3} |f_{4}\rangle+\langle f_{6} |f_{4}\rangle+\langle f_{1} |f_{5}\rangle+\langle f_{4} |f_{5}\rangle+\langle f_{7} |f_{5}\rangle\\
& \quad +\langle f_{2} |f_{6}\rangle+\langle f_{5} |f_{6}\rangle+\langle f_{8} |f_{6}\rangle+\langle f_{0} |f_{7}\rangle+\langle f_{3} |f_{7}\rangle+\langle f_{6} |f_{7}\rangle+\langle f_{1} |f_{8}\rangle+\langle f_{4} |f_{8}\rangle+\langle f_{7} |f_{8}\rangle)\\
& \quad +\frac{1}{9}e^{\frac{-2i\pi}{3}}Re(\langle f_{1}| f_{0}\rangle+\langle f_{4} |f_{0}\rangle+\langle f_{7} |f_{0}\rangle+\langle f_{2} |f_{1}\rangle+\langle f_{5}| f_{1}\rangle+\langle f_{8} |f_{1}\rangle+\langle f_{0} |f_{2}\rangle+\langle f_{3} |f_{2}\rangle \\
& \quad +\langle f_{6} f_{2}\rangle+\langle f_{1} |f_{3}\rangle+\langle f_{4} |f_{3}\rangle+\langle f_{7} |f_{3}\rangle+\langle f_{2}| f_{4}\rangle+\langle f_{5} |f_{4}\rangle+\langle f_{8} |f_{4}\rangle+\langle f_{0} |f_{5}\rangle+\langle f_{3} |f_{5}\rangle+\langle f_{6} |f_{5}\rangle\\
& \quad +\langle f_{1} |f_{6}\rangle+\langle f_{4} |f_{6}\rangle+\langle f_{7} |f_{6}\rangle+\langle f_{2} |f_{7}\rangle+\langle f_{5} |f_{7}\rangle+\langle f_{8} |f_{7}\rangle+\langle f_{0} |f_{8}\rangle+\langle f_{3} |f_{8}\rangle+\langle f_{6} |f_{8}\rangle)    
    \end{split}
\end{equation}

\begin{equation}
    \begin{split}
p_{1^{\prime\prime}0^{\prime\prime}}=&\langle h_{3} |h_{3}\rangle\\=&\frac{1}{3}+\frac{1}{9}Re(\langle f_{1} |f_{0}\rangle+\langle f_{2} |f_{0}\rangle+\langle f_{0} |f_{1}\rangle+\langle f_{2} |f_{1}\rangle+\langle f_{0} |f_{2}\rangle+\langle f_{1} |f_{2}\rangle+\langle f_{4} |f_{3}\rangle+\langle f_{5} |f_{3}\rangle\\
& \quad +\langle f_{3} |f_{4}\rangle+\langle f_{5}|f_{4}\rangle+\langle f_{3} |f_{5}\rangle+\langle f_{4} |f_{5}\rangle+\langle f_{7} |f_{6}\rangle+\langle f_{8} |f_{6}\rangle+\langle f_{6} |f_{7}\rangle+\langle f_{8} |f_{7}\rangle+\langle f_{6} |f_{8}\rangle+\langle f_{7} |f_{8}\rangle)\\
& \quad +\frac{1}{9}e^{\frac{-2i\pi}{3}}Re(\langle f_{4} |f_{0}\rangle+\langle f_{5} |f_{0}\rangle+\langle f_{3} |f_{1}\rangle+\langle f_{5} |f_{1}\rangle+\langle f_{3}| f_{2}\rangle+\langle f_{4} |f_{2}\rangle+\langle f_{7} |f_{3}\rangle+\langle f_{8} |f_{3}\rangle \\
& \quad +\langle f_{6} |f_{4}\rangle+\langle f_{8} |f_{4}\rangle+\langle f_{6} |f_{5}\rangle+\langle f_{7} |f_{5}\rangle+\langle f_{1} |f_{6}\rangle+\langle f_{2} |f_{6}\rangle+\langle f_{0} |f_{7}\rangle+\langle f_{2} |f_{7}\rangle+\langle f_{0} |f_{8}\rangle+\langle f_{1} |f_{8}\rangle)\\
& \quad  +\frac{1}{9}e^{\frac{2i\pi}{3}}Re(\langle f_{7} |f_{0}\rangle+\langle f_{8} |f_{0}\rangle+\langle f_{6} |f_{1}\rangle+\langle f_{8} |f_{1}\rangle+\langle f_{6} |f_{2}\rangle+\langle f_{7} |f_{2}\rangle+\langle f_{1} |f_{3}\rangle+\langle f_{2} |f_{3}\rangle+\langle f_{0} |f_{4}\rangle \\
& \quad +\langle f_{2} |f_{4}\rangle+\langle f_{0} |f_{5}\rangle+\langle f_{1} |f_{5}\rangle+\langle f_{4} |f_{6}\rangle+\langle f_{5} |f_{6}\rangle+\langle f_{3} |f_{7}\rangle+\langle f_{5} |f_{7}\rangle+\langle f_{4} |f_{8}\rangle+\langle f_{3} |f_{8}\rangle
    \end{split}
\end{equation}

\begin{equation}
    \begin{split}
    p_{1^{\prime\prime}2^{\prime\prime}} =& \langle h_{5}|h_{5}\rangle \\=&\frac{1}{3}+\frac{1}{9}Re(\langle f_{5}|f_ {0}\rangle+\langle f_{7}|f_{0}\rangle +\langle f_{3}|f_{1} \rangle+ \langle f_{8}|f_{1}\rangle+\langle f_{4}|f_{2} \rangle+\langle f_{6}|f_{2} \rangle+\langle f_{1}|f_{3}\rangle+\langle f_{8}|f_{3} \rangle+\langle f_{2}|f_{4}\rangle+\langle f_{6}|f_{4} \rangle\\
& \quad +\langle f_{0}|f_{5}\rangle+\langle f_{7}|f_{5} \rangle+\langle f_{2}|f_{6}\rangle+\langle f_{4}|f_{6}\rangle+\langle f_{0}|f_{7}\rangle+\langle f_{5}|f_{7}\rangle+\langle f_{1}|f_{8}\rangle+\langle f_{3}|f_{8} \rangle) \\
& \quad  +\frac {1}{9}e^{\frac{-2 i\pi}{3}}Re(\langle f_{1}|f_{0}\rangle+\langle f_{8}|f_{0}\rangle+\langle f_{2}|f_{1} \rangle + \langle f_{6}|f_{1}\rangle+\langle f_{0}|f_{2}\rangle +\langle f_{7}|f_{2} \rangle+\langle f_{2} |f_{3} \rangle+\langle f_{4}|f_{3} \rangle \\
& \quad + \langle f_{0}|f_{4} \rangle + \langle f_{5}|f_{4} \rangle + \langle f_{1}|f_{5} \rangle + \langle f_{3}|f_{5} \rangle + \langle f_{5}|f_{6} \rangle + \langle f_{7}|f_{6} \rangle + \langle f_{3}|f_{7} \rangle + \langle f_{8}|f_{7} \rangle + \langle f_ {4}|f_{8} \rangle +\langle f_{6}|f_{8} \rangle) \\
&  \quad  + \frac {1} {9} e^{\frac {2 i\pi} {3}}Re(\langle f_ {2} |f_{0} \rangle + \langle f_{4}|f_{0} \rangle + \langle f_{0}|f_{1} \rangle + \langle f_{5}|f_{1} \rangle + \langle f_{1}|f_{2} \rangle + \langle f_{3}|f_{2} \rangle + \langle f_{5}|f_{3} \rangle + \langle f_{7}|f_{3} \rangle + \langle f_{3}|f_{4}\rangle \\
& \quad +\langle f_{8}|f_{4}\rangle + \langle f_{4}|f_{5} \rangle+\langle f_{6}|f_{5} \rangle + \langle f_{1}|f_{6} \rangle + \langle f_{8}|f_{6} \rangle + \langle f_{2}|f_{7} \rangle + \langle f_{6}|f_{7} \rangle + \langle f_{0}|f_{8} \rangle + \langle f_{7}|f_{8}\rangle)    
    \end{split}
\end{equation}

\begin{equation}
\begin{split}
 p_ {2^{\prime\prime }0^{\prime \prime }}= & \langle h_{6}|h_{6}\rangle\\
 = &\quad  \frac{1}{3}+\frac{1}{9}Re(\langle f_{2}|f_{1}\rangle +\langle f_{5}|f_{3}\rangle +\langle f_{0}|f_{1} \rangle +\langle f_{1}|f_{2}\rangle +\langle f_{4}|f_{3}\rangle  +\langle f_{0}|f_{2}\rangle +\langle f_{2}|f_{0}\rangle +\langle f_{5}|f_{4}\rangle \\
 &\quad +\langle f_{3}|f_{4} \rangle +\langle f_{1}|f_{0} \rangle +\langle f_{4}|f_{5}\rangle +\langle f_{3}|f_{5}\rangle +\langle f_{7}|f_{6}\rangle +\langle f_{8}|f_{6}\rangle +\langle f_{6}|f_{7}\rangle +\langle f_{6}|f_{8}\rangle +\langle f_{8}| f_{7}\rangle +\langle f_{7}|f_{8}\rangle )\\
 &\quad + \frac{1}{9}e^{\frac{2 i\pi }{3}}Re(\langle f_{1}|f_{6}\rangle+\langle f_{2}|f_{6}\rangle +\langle f_{4}|f_{0}\rangle +\langle f_{5}|f_{0}\rangle +\langle f_{1}|f_{8}\rangle +\langle f_{2}|f_{7}\rangle +\langle f_{3}|f_{1}\rangle +\langle f_{3}|f_{2}\rangle \\
 &\quad +\langle f_{4}|f_{2}\rangle +\langle f_{5}|f_{1}\rangle +\langle f_{4}|f_{8}\rangle +\langle f_{5}|f_{7}\rangle +\langle f_{6}|f_{4}\rangle +\langle f_{6}|f_{5}\rangle +\langle f_{0}|f_{7}\rangle +\langle f_{7}|f_{3}\rangle +\langle f_{0}|f_{8}\rangle +\langle f_{8}|f_{3}\rangle )\\ 
 &\quad + \frac{1}{9} e^{\frac{-2 i\pi }{3}}Re( \langle f_{7}|f_{0}\rangle +\langle f_{8}|f_{0}\rangle +\langle f_{6}|f_{1}\rangle +\langle f_{1}|f_{3}\rangle +\langle f_{6}|f_{2}\rangle +\langle f_{2}|f_{3}\rangle +\langle f_{3}|f_{7}\rangle +\langle f_{3}|f_{8}\rangle +\langle f_{0}|f_{4}\rangle +\langle f_{2}|f_{4}\rangle \\
&\quad +\langle f_{0}|f_{5}\rangle +\langle f_{1}|f_{5}\rangle +\langle f_{4}|f_{6}\rangle +\langle f_{5}|f_{6}\rangle +\langle f_{7}|f_{2}\rangle +\langle f_{5}|f_{7}\rangle +\langle f_{8}|f_{1}\rangle +\langle f_{4}|f_{8}\rangle) 
\end{split}
\end{equation}

\begin{equation}
\label{pijprimeprime2}
    \begin{split}
p_ {2^{\prime\prime}1^{\prime\prime}} =& \langle h_ {7}|h_ {7} \rangle\\
=&\frac {1}{3}+\frac{1}{9}Re(\langle f_{5}|f_{0} \rangle + \langle f_{7}|f_{0} \rangle + \langle f_{3}|f_{1} \rangle+\langle f_{8}|f_{1}\rangle+\langle f_{4}|f_{2} \rangle + \langle f_{6}|f_{2} \rangle + \langle f_ {1} | f_ {3} \rangle+\langle f_{8}|f_{3} \rangle + \langle f_{2}|f_{4} \rangle+\langle f_{6}|f_{4} \rangle \\
& \quad + \langle f_{0}|f_{5} \rangle + \langle f_{7}|f_{5} \rangle + \langle f_{2}|f_{6} \rangle + \langle f_{4}|f_{6} \rangle + \langle f_{0}|f_{7} \rangle+ \langle f_{5}|f_{7} \rangle+ \langle f_{1}|f_{8} \rangle + \langle f_{3}|f_{8} \rangle  \\
& \quad  + \frac {1} {9} e^{\frac {2 i\pi} {3}}Re(\langle f_{1}|f_{0} \rangle + \langle f_{8}|f_{0} \rangle+\langle f_{2}|f_{1} \rangle + \langle f_{6}|f_{1} \rangle+\langle f_{0}|f_{2} \rangle+ \langle f_{7}|f_{2} \rangle + \langle f_{2}|f_{3}\rangle + \langle f_{4}|f_{3} \rangle \\
& \quad  + \langle f_{0}|f_{4} \rangle +\langle f_{5}|f_{4} \rangle + \langle f_{1}|f_{5} \rangle + \langle f_{3}|f_{5} \rangle+ \langle f_{5}|f_{6} \rangle + \langle f_{7}|f_{6} \rangle + \langle f_{3}|f_{7}\rangle + \langle f_{8}|f_{7} \rangle + \langle f_ {6}| f_ {8} \rangle + \langle f_ {4}| f_ {8} \rangle) \\
& \quad  + \frac {1} {9} e^{\frac {-2 i\pi} {3}}Re(\langle f_ {2} | f_ {0} \rangle + \langle f_ {4} | f_ {0} \rangle + \langle f_ {0} | f_ {1} \rangle + \langle f_ {5} | f_ {1} \rangle + \langle f_ {1} | f_ {2} \rangle + \langle f_ {3} | f_ {2} \rangle + \langle f_ {5} | f_ {3} \rangle + \langle f_ {7} | f_ {3} \rangle + \langle f_ {3} | f_ {4} \rangle \\ 
& \quad  + \langle f_ {8} |f_ {4} \rangle + \langle f_ {4} | f_ {5} \rangle + \langle f_ {6} | f_ {5} \rangle + \langle f_ {1} | f_ {6} \rangle + \langle f_ {8} | f_ {6} \rangle + \langle f_ {2} | f_ {7} \rangle + \langle f_ {6} | f_ {7} \rangle + \langle f_ {0} | f_ {8} \rangle + \langle f_ {7} | f_ {8} \rangle)    
    \end{split}
\end{equation}

Using the Cauchy-Schwarz inequality, we may bound the quantity $X_{\phi_2}=Re(\langle e^0_{0,0}|e^1_{1,4}\rangle)+Re(\langle e^0_{0,0}|e^2_{2,8}\rangle)+Re(\langle e^1_{1,4}|e^2_{2,8}\rangle)$ as follows
\begin{equation}
\label{upperbound2}
\begin{split}
    X_{\phi_2}  \geq & 3-\frac{3}{2}(p_{0^{\prime\prime}1^{\prime\prime}}+ p_{0^{\prime\prime}2^{\prime\prime}}+ p_{1^{\prime\prime}0^{\prime\prime}}+ p_{1^{\prime\prime}2^{\prime\prime}}+ p_{2^{\prime\prime}0^{\prime\prime}}+ p_{2^{\prime\prime}1^{\prime\prime}} \\
& \quad -(\sqrt{p_{001}p_{102}} + \sqrt{p_{011}p_{102}} + \sqrt{p_{021}p_{102}} + \sqrt{p_{001}p_{112}} + \sqrt{p_{011}p_{112}} + \sqrt{p_{021}p_{112}} \\
& \quad +  \sqrt{p_{001}p_{122}} + \sqrt{p_{011}p_{122}} + \sqrt{p_{021}p_{122}} + \sqrt{p_{001}p_{200}} + \sqrt{p_{011}p_{200}} + \sqrt{p_{021}p_{200}} \\
& \quad +  \sqrt{p_{001}p_{210}} + \sqrt{p_{011}p_{210}} + \sqrt{p_{021}p_{210}} + \sqrt{p_{001}p_{220}} + \sqrt{p_{011}p_{220}} + \sqrt{p_{021}p_{220}} \\
& \quad +  \sqrt{p_{002}p_{100}} + \sqrt{p_{012}p_{100}} + \sqrt{p_{022}p_{100}} + \sqrt{p_{002}p_{110}} + \sqrt{p_{012}p_{110}} + \sqrt{p_{022}p_{110}} \\ 
& \quad +  \sqrt{p_{002}p_{120}} + \sqrt{p_{012}p_{120}} + \sqrt{p_{022}p_{120}} + \sqrt{p_{002}p_{201}} + \sqrt{p_{012}p_{201}} + \sqrt{p_{022}p_{201}} \\
& \quad +  \sqrt{p_{002}p_{211}} + \sqrt{p_{012}p_{211}} + \sqrt{p_{022}p_{211}} + \sqrt{p_{002}p_{221}} + \sqrt{p_{012}p_{221}} + \sqrt{p_{022}p_{221}} \\
& \quad +  \sqrt{p_{100}p_{201}} + \sqrt{p_{110}p_{201}} + \sqrt{p_{120}p_{201}} + \sqrt{p_{100}p_{211}} + \sqrt{p_{110}p_{211}} + \sqrt{p_{120}p_{211}} \\
& \quad +  \sqrt{p_{100}p_{221}} + \sqrt{p_{110}p_{221}} + \sqrt{p_{120}p_{221}} + \sqrt{p_{102}p_{200}} + \sqrt{p_{112}p_{200}} + \sqrt{p_{122}p_{200}} \\
& \quad +  \sqrt{p_{102}p_{210}} + \sqrt{p_{112}p_{210}} + \sqrt{p_{122}p_{210}} + \sqrt{p_{102}p_{220}} + \sqrt{p_{112}p_{220}} + \sqrt{p_{122}p_{220}})\\
&-(\sqrt {p_{000} p_{101}}+\sqrt {p_{010} p_{101}}+\sqrt {p_{020} p_{101}}+\sqrt {p_{010} p_{111}}+\sqrt {p_{020} p_{111}} +\sqrt {p_{000} p_{121}}\\
&+\sqrt {p_{010} p_{121}}+\sqrt {p_{020} p_{121}}+\sqrt {p_{000} p_{202}}+\sqrt {p_{010} p_{202}} +\sqrt {p_{020} p_{202}}+\sqrt {p_{000} p_{212}}\\
&+\sqrt {p_{010} p_{212}} +\sqrt {p_{020} p_{212}} + \sqrt {p_{010}p_{222}}+\sqrt {p_{020} p_{222}}+\sqrt {p_{101} p_{202}}+\sqrt {p_{111} p_{202}} \\
&+\sqrt {p_{121} p_{202}}+\sqrt {p_{101} p_{212}} +\sqrt {p_{111} p_{212}} + \sqrt {p_{121} p_{212}} +\sqrt {p_{101} p_{222}} + \sqrt {p_{121} p_{222}}).
\end{split}
\end{equation}

\end{document}